%% file: paper.tex
\documentclass[iop, apj]{emulateapj}

\usepackage{xspace}
\usepackage{amsmath}
\usepackage{framed} 
\usepackage{txfonts}
\usepackage{epstopdf} 
\usepackage{color}
\usepackage{rotating}
\usepackage{natbib}
\usepackage{ulem}
\usepackage{hyperref}

\hypersetup{
  colorlinks   = true, %Colours links instead of ugly boxes
  urlcolor     = black, %Colour for external hyperlinks
  linkcolor    = black, %Colour of internal links
  citecolor   = black %Colour of citations
}

\journalinfo{The Astrophysical Journal, {\rm 766:25 (15pp), 2013 March 20}}
\submitted{Received 2012 July 3; accepted 2013 January 30; published 2013 March 4}

\input{commands}

\input{macros}

\shorttitle{The pseudo-evolution of halo mass}
\shortauthors{Diemer, More \& Kravtsov}

\begin{document}
\setlength{\hbadness}{10000}

%%%%%%%%%%%%%%%%%%%%%%%%%%%%%%%%%%%%%%%%%%%%%%%%%%%%%%%%%%%%%%%%%%%%%%%%%%
% TITLE ETC
%%%%%%%%%%%%%%%%%%%%%%%%%%%%%%%%%%%%%%%%%%%%%%%%%%%%%%%%%%%%%%%%%%%%%%%%%%

\title{The pseudo-evolution of halo mass}

\author{Benedikt Diemer \altaffilmark{1,2}, Surhud More \altaffilmark{2,3}, \& Andrey V. Kravtsov\altaffilmark{1,2,3}}

\affil{ 
$^1$ Department of Astronomy \& Astrophysics, The University of Chicago, Chicago, IL 60637 USA; {\tt bdiemer@oddjob.uchicago.edu} \\
$^2$ Kavli Institute for Cosmological Physics, The University of Chicago, Chicago, IL 60637 USA \\
$^3$ Enrico Fermi Institute, The University of Chicago, Chicago, IL 60637 USA}

%%%%%%%%%%%%%%%%%%%%%%%%%%%%%%%%%%%%%%%%%%%%%%%%%%%%%%%%%%%%%%%%%%%%%%%%%%
% ABSTRACT
%%%%%%%%%%%%%%%%%%%%%%%%%%%%%%%%%%%%%%%%%%%%%%%%%%%%%%%%%%%%%%%%%%%%%%%%%%

\begin{abstract}
A dark matter halo is commonly defined as a spherical overdensity of matter with respect to a reference density, such as the critical density or the mean matter density of the universe. Such definitions can lead to a spurious {\it pseudo-evolution} of halo mass simply due to redshift evolution of the reference density, even if its physical density profile remains constant over time.  We estimate the amount of such pseudo-evolution of mass between $z=1$ and $0$ for halos identified in a large N-body simulation, and show that it  accounts for almost the entire mass evolution of the majority of halos with $M_{200\bar\rho} \simlt 10^{12} \msunh$ and can be a significant fraction of the apparent mass growth even for cluster-sized halos. We estimate the magnitude of the pseudo-evolution assuming that halo density profiles remain static in physical coordinates, and show that this simple model predicts the pseudo-evolution of halos identified in numerical simulations to good accuracy, albeit with significant scatter. We discuss the impact of pseudo-evolution on the evolution of the halo mass function and show that the non-evolution of the low-mass end of the halo mass function is the result of a fortuitous cancellation between pseudo-evolution and the absorption of small halos into larger hosts. We also show that the evolution of the low mass end of the concentration-mass relation observed in simulations is almost entirely due to the pseudo-evolution of mass. Finally, we discuss the implications of our results for the interpretation of the evolution of various scaling relations between the observable properties of galaxies and galaxy clusters and their halo masses.
\end{abstract}


%%%%%%%%%%%%%%%%%%%%%%%%%%%%%%%%%%%%%%%%%%%%%%%%%%%%%%%%%%%%%%%%%%%%%%%%%%
% INTRODUCTION
%%%%%%%%%%%%%%%%%%%%%%%%%%%%%%%%%%%%%%%%%%%%%%%%%%%%%%%%%%%%%%%%%%%%%%%%%%

\section{Introduction}
\label{sec:intro}

In a cold dark matter cosmological scenario \citep[see, e.g.,][]{peebles_84_cdm, davis_85_clustering}, the drama of galaxy formation unfolds at the virialized peaks of the density field, or {\it halos}. Although galaxies themselves are highly diverse, several of their properties exhibit remarkable regularity and can be expressed as galaxy scaling relations. In particular, the stellar mass-halo mass relation and the luminosity-halo mass relation of central galaxies constrain important aspects of galaxy formation and have been studied via a variety of probes such as satellite kinematics \citep{prada_03_sk, conroy_07_sk, more_09_sk, more_11_sk}, galaxy-galaxy weak lensing \citep{seljak_00_wl, mckay_01_wl, mandelbaum_06_shmr, parker_07_wl, schulz_10_wl}, the abundance of galaxies and their clustering \citep{yang_03_hod, zehavi_04_hod, tinker_05_hod, zehavi_05_hod, skibba_06_hod, vandenbosch_07_hod, brown_08_hod, conroy_09_shmr, moster_10_shmr, moster_13_shmr, behroozi_10_shmr, yang_12_dm_galaxy}, or a combination of the above probes \citep{yoo_06_clust_wl, cacciato_09_clust_wl, leauthaud_12_shmr, more_12_galaxy_dm}. In order to understand the formation and evolution of galaxies, it is crucial to interpret the evolution of these scaling relations, which in turn requires a solid understanding of the evolution of halo masses with cosmic time.

Analogously, the largest halos in the universe host clusters of galaxies, which themselves serve as laboratories for galaxy formation. The observable properties of clusters, such as X-ray temperature, entropy profile, the mass of the intracluster gas, or their evolution with redshift, are often described using a self-similar model \citep[][see also \citealt{kravtsov_12_cluster_review} for a review]{kaiser_86_clusters}. This model provides predictions for the scaling relations between halo mass and the observable properties of clusters. Large observational campaigns have been undertaken in the past \citep{vikhlinin_06_clusters, bohringer_07_rexcess, mantz_10_clusters} and are also currently under way \citep[e.g.][]{benson_11_spt_xray} to calibrate these scaling relations since they are necessary to obtain cosmological constraints from the observed abundance of clusters and its redshift evolution \citep[e.g.,][see \citealt{allen_11_clusterreview} for a recent review]{vikhlinin_09_xray}. However, such observational campaigns must be supplemented by sound theoretical models for the evolution of the scaling relations, which have still not been fully developed \citep[see, e.g., a recent analysis by][]{lin_12_baryoncontent}. 

When quoting the scaling relations between halo mass and galaxy (or galaxy cluster) properties, observers inevitably adopt a specific definition for the boundaries of  halos, often based on the extent of their observations. However, numerical simulations show that dark matter halos exhibit smooth density profiles without well-defined boundaries, which makes the definition of the halo boundary and the associated halo mass ambiguous. The mass definition often used in the literature corresponds to the mass within a spherical boundary that encloses a given overdensity, $\Delta(z)$, with respect to a reference density, $\rho_{\rm ref}(z)$ \citep[see, e.g.,][]{cole_96_halostructure}. This spherical overdensity (SO) halo mass, $M_{\Delta}(z)$, and radius, $R_{\Delta}(z)$, are thus related via the following equation:
\begin{eqnarray}
M_{\Delta}(z)=\frac{4}{3}\pi R_{\Delta}^3(z) \Delta(z)\,\rho_{\rm ref}(z)\,.
\label{eq:mdelta1}
\end{eqnarray}
The most common choices of reference density are either the critical density, $\rho_\rmc$, or the mean matter density, $\bar{\rho}$, of the universe at a given cosmic epoch. The parameter $\Delta$ can be chosen arbitrarily, but certain values such as $\Delta=180$ can be justified with the spherical top hat collapse model for an Einstein-de Sitter cosmology \citep{gunn_72_sphericalcollapse}. The spherical collapse model has also been generalized to cosmological models which include a cosmological constant or non-zero curvature \citep{lahav_91_lambda_clusters, lacey_93_collapse, eke_96_clustercosmo}. 

The fundamental issue with the mass definition of Equation (\ref{eq:mdelta1}) is that the reference density evolves with cosmic time, leading to an evolution in halo mass even if the physical density profile of the halo is constant. For the remainder of this paper, we shall call the evolution of halo mass due to changing reference density {\it pseudo-evolution} because it is due solely to the mass definition and not to any actual physical mass evolution caused by the accretion of new material. Note that the actual evolution of SO mass, which we shall call {\it mass evolution}, is a combination of the physical evolution due to the accretion of matter and pseudo-evolution. 

The fact that the evolution of the SO mass may not correspond to any actual physical evolution of mass has been pointed out before. \citet{diemand_07_haloevolution} analyzed the accretion history of the Milky Way sized Via Lactea halo and found no significant physical growth after $z = 1$, even though the virial mass of the halo increased significantly. \citet{prada_06_outerregions} studied the outer regions of collapsed halos at $z = 0$ and found no systematic infall for halos with masses lower than $5\times 10^{12} \msunh$. In a follow-up study, \citet{cuesta_08_infall} demonstrated a lack of physical accretion onto galaxy mass halos, and proposed an alternative mass definition that aims to include all mass bound to a halo \citep[see also][]{anderhalden_11_totalmass}. Although such a mass definition may be more physical and closer to the meaning of mass in analytical models of halo collapse and evolution, its observational analog is very difficult or even impossible to measure for real systems. Thus, the SO mass is most often used in observations, and a proper interpretation of observational results should take into account the pseudo-evolution inherent in this mass definition. For the case of cluster scaling relations, \citet{kravtsov_12_cluster_review} argued that part of their evolution is due to pseudo-evolution.

In this paper, we seek to quantify the pseudo-evolution of the SO  mass accretion history (MAH) of halos due to an evolving reference density.  Much work has been invested into quantifying MAHs, but the contributions from physical accretion and pseudo-evolution are generally not distinguished \citep[see, e.g.,][]{wechsler_02_halo_assembly, vandenbosch_02_mah, zhao_03_mah, miller_06_mah, zhao_09_mah}.  We will show that the contribution from pseudo-evolution to MAH depends upon the time evolution of halo density profile, some aspects of which have been investigated previously. For example, it has been demonstrated that the scale radius $r_s$ and scale density $\rho_s$ of galactic-sized halos do not evolve significantly after $z=1$ \citep[see, e.g.,][]{bullock_01_halo_profiles, wechsler_02_halo_assembly, zhao_03_mah}. However, as we demonstrate in Section \ref{subsec:nonstatic}, determining the amount of pseudo-evolution requires knowledge of the evolution of the outer regions of the density profile (around the boundary $R_{\Delta}$). The non-evolution of the outer regions for galactic-sized halos has been presented before \citep{diemand_07_haloevolution, cuesta_08_infall}, but for masses limited to Milky Way sized halos. We extend the results from these studies by comparing density profiles for a wide range of halo masses at different redshifts. We quantify the mean and the scatter of the contributions from physical accretion and pseudo-evolution to the mass evolution histories. Furthermore, we investigate the impact of pseudo-evolution on the evolution of the halo mass function, and show that pseudo-evolution can account for the majority of the observed evolution in the concentration-mass relation since $z=1$. 

This paper is structured as follows. In Section \ref{sec:theory}, we derive mathematical definitions for the pseudo-evolution in the cases of static and evolving halo density profiles. In Section \ref{sec:profiles}, we quantify the pseudo-evolution of actual halos in cosmological simulations. We discuss caveats and implications of our results as well as directions for future work in Section \ref{sec:discussion}, and give a summary of our results in Section \ref{sec:conclusion}. Throughout this paper, we denote overdensities as $\Delta_\rmc$ if they are defined relative to $\rho_\rmc$, and $\Delta_\rmm$ if they are defined relative to $\bar{\rho}$. We also use $\Delta_{\rm vir}$ to denote the redshift and cosmology-dependent virial overdensity predicted by the spherical collapse model, which corresponds to $\Delta_{\rm vir}\approx 358$ at $z=0$ and $\Delta_{\rm vir}\approx 180$ at $z>2$ with respect to the mean background density for the concordance fiducial cosmology used in this paper \citep[e.g.,][]{bryan_98_virial}. All densities and radii are expressed in physical units, unless stated otherwise.

%%%%%%%%%%%%%%%%%%%%%%%%%%%%%%%%%%%%%%%%%%%%%%%%%%%%%%%%%%%%%%%%%%%%%%%%%%
% THE STATIC HALO MODEL
%%%%%%%%%%%%%%%%%%%%%%%%%%%%%%%%%%%%%%%%%%%%%%%%%%%%%%%%%%%%%%%%%%%%%%%%%%

\section{Theoretical considerations}
\label{sec:theory}

Before quantifying the amount of pseudo-evolution in simulated halos, we investigate some simplified scenarios of halo growth. In Section \ref{subsec:staticmodel}, we quantify pseudo-evolution in the case of static density profiles. We describe a simple analytical model based on the Navarro-Frenk-White (NFW) density profiles \citep{navarro_97_nfw} to gauge the contribution of pseudo-evolution to the total MAH. In Section \ref{subsec:nonstatic}, we allow the density profile to increase or decrease monotonically, and derive estimators for the contribution of pseudo-evolution to the total evolution. We finish with a discussion of the redshift range most suitable to investigate pseudo-evolution in Section \ref{subsec:zrange}.

\subsection{Pseudo-evolution in Static Halos}
\label{subsec:staticmodel}

\begin{figure}
\centering
\includegraphics[trim = 0mm 5mm 100mm 100mm, clip, scale=0.65]{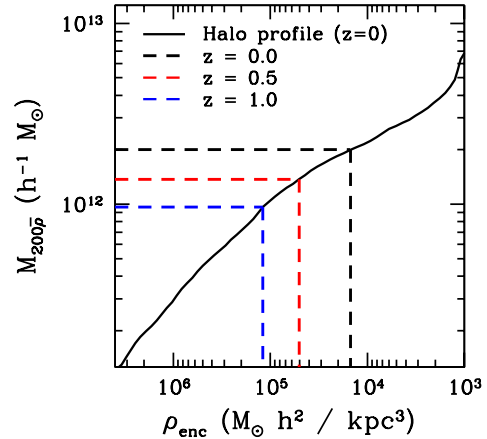} 
\caption{Visualization of the static halo model. The solid line shows the spherical mass profile as a function of enclosed density ($M / V$) for a halo of mass $\mmean=2\times 10^{12} \msunh$ from the Bolshoi simulation at $z = 0$. The x-axis is reversed, so that the left side of the plot corresponds to the high-density center of the halo, and the right to the low-density outskirts. The vertical dashed lines indicate the spherical overdensity $200 \bar{\rho}$ at redshifts $0$, $0.5$ and $1$, and the horizontal dashed lines mark the corresponding halo mass $\mmean$. As the reference density evolves, $\bar{\rho} \propto (1+z)^3$, the halo density threshold increases with redshift, and $\mmean$ decreases. Even if the physical mass distribution of this halo was kept fixed between $z=1$ and $z=0$, its mass $\mmean$ would undergo a {\it pseudo-evolution} from $9.6 \times 10^{11} \msunh$ to $2 \times 10^{12} \msunh$.}
\label{fig:static}
\end{figure}

\begin{figure}
\centering
\includegraphics[trim = 5mm 7mm 88mm 38mm, clip, scale=0.73]{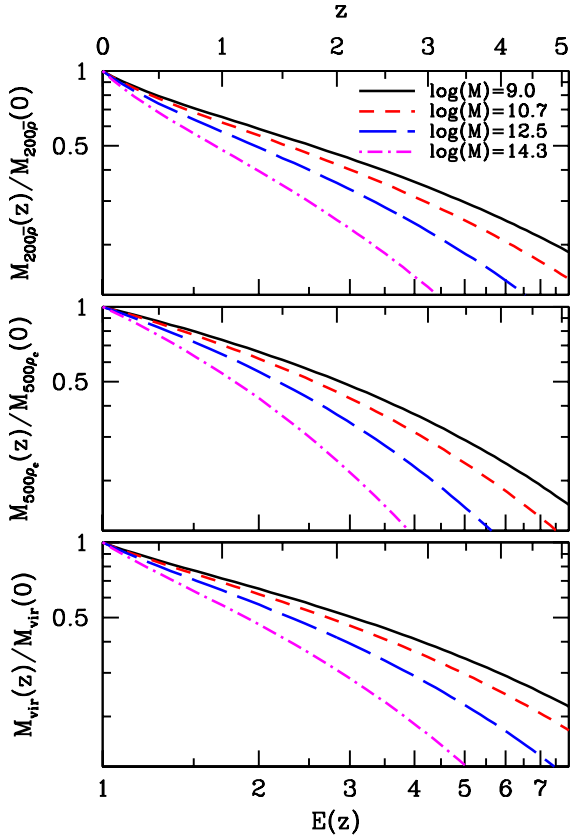} 
\caption{Predictions of the static halo model. Lines show the {\it pseudo-evolution} of halo mass due to changing reference density as a function of $E(z)$ relative to the halo mass at $z=0$. The labels indicate the halo mass in $\log(\msunh)$. From top to bottom, the panels show the evolution of $\mmean$, $\mcrit$ and $\mvir$, where $\bar{\rho}$ and $c$ indicate the average and critical densities of the universe, respectively, and $\mvir$ corresponds to an evolving overdensity according to \citet{bryan_98_virial} . For $z<0.5$, the pseudo-evolution is largest for $\mmean$, but the overall trend is the same for all three mass definitions.}
\label{fig:analytical_prediction}
\end{figure}

Let us consider a density peak in the universe around which the matter density profile in physical units has not evolved since a given initial redshift, $z_\rmi$. As shown in Figure \ref{fig:static}, the halo mass associated with this density peak will change purely due to the evolution of the reference density used to define its boundary. This evolution in mass can be quantified using the density profile of the halo at redshift $z_\rmi$. Let us assume that the density distribution around this density peak is described by the universal density profile given by
\begin{eqnarray}
\rho(r, z_\rmi)=\frac{\rho_\rms}{\left(r/r_\rms\right) \left(1+r/r_\rms\right)^2} \,,
\label{eq:nfw}
\end{eqnarray}
which has been found to be a reasonable approximation of the typical density profiles around density peaks in cold dark matter cosmologies \citep[][hereafter NFW]{navarro_97_nfw}. The scale radius $r_\rms$ and the halo radius $R_{\Delta}$  are related by the concentration parameter $c_{\Delta}=R_{\Delta}/r_\rms$. Under our assumption that the density profile around this peak does not evolve and that profile of Equation (\ref{eq:nfw}) is a good description of the actual profile at the radii of interest, the halo mass at any redshift $z$ can be expressed in terms of the characteristic density $\rho_\rms$ and $r_\rms$, by integrating this {\it static} density profile within the halo radius $R_{\Delta}(z)$, such that
\begin{equation}
M_\Delta(z)=\int_{0}^{R_\Delta(z)} \rho(r) 4 \pi r^2 dr = \rho_s 4 \pi
r_s^3 \mu[c_\Delta(z)]\,,
\label{eq:mdelta2}
\end{equation}
where the function $\mu[x]$ is given by
\begin{eqnarray}
\label{eq:mu}
\mu[x]=\ln(1+x)-\frac{x}{1+x}\,.
\end{eqnarray}
Equating the right hand sides of Equations (\ref{eq:mdelta1}) and (\ref{eq:mdelta2}), we obtain a relation between the concentration parameter of the halo at redshift $z$ and the concentration parameter at redshift $z_\rmi$,
\begin{eqnarray}
\frac{c_{\Delta}(z)^3}{\mu[c_\Delta(z)]} &=&
\frac{3\rho_s}{\Delta(z)\rho_{\rm ref}(z)} \,\\
&=&
\frac{c_{\Delta}(z_\rmi)^3}{\mu[c_\Delta(z_\rmi)]}
\left[\frac{\Delta(z_\rmi)\rho_{\rm
ref}(z_\rmi)}{\Delta(z)\rho_{\rm ref}(z)}\right]\,.
\label{eq:cevol}
\end{eqnarray}
This relation can in turn be used to find the evolution of halo mass according to the equation
\begin{eqnarray}
M_{\Delta}(z)=M_{\Delta}(z_\rmi)\frac{\mu[c_{\Delta}(z)]}{\mu[c_{\Delta}(z_\rmi)]}\,.
\label{eq:mevol}
\end{eqnarray}

As examples, we consider three commonly used definitions of halo mass in the literature, (1) $\Delta_m(z)=200$ as in studies of the halo occupation distribution of galaxies, (2) $\Delta_\rmc(z)=500$ as in studies involving galaxy cluster observations, and (3) $\Delta_\rmc(z)=\Delta_{\rm vir}$. Without loss of generality, we use $z_\rmi=0$ to define the static density profile in physical units. We consider the concentration-mass relation at $z=0$ given by \citet[][hereafter Z09]{zhao_09_mah}, and use Equations (\ref{eq:cevol}) and (\ref{eq:mevol}) to obtain the mass evolution due to pseudo-evolution. 

Figure \ref{fig:analytical_prediction} shows the mass evolution histories for halos of different masses as predicted by our static halo model. The different panels correspond to the three commonly used overdensity definitions. Each panel shows the pseudo-evolution of mass from $z=0$ to $z=5$, normalized to the halo's mass at $z=0$, as a function of the expansion rate in units of the Hubble constant, $E(z)$. Assuming a flat $\Lambda$CDM cosmology, $E(z)$ is defined as usual,
\begin{equation}
E(z)=\sqrt{\Omega_{\Lambda}+\Omega_m (1+z)^3}.
\end{equation}
As expected, more massive halos undergo a larger evolution due to the lower values of their concentrations. Regardless of the exact mass definition, the fractional change in halo mass due to pseudo-evolution can be as large as $\sim0.5$ by $z=1$. The shape of the mass evolution history with redshift is not only a function of halo mass, but also depends upon the exact mass definition. Its functional form is better approximated by a power law of $E(z)$ than $(1+z)$, but still shows deviations from an exact power law behavior.

\subsection{Pseudo-evolution in Physically Accreting Halos}
\label{subsec:nonstatic}

\begin{figure*}
\centering
\includegraphics[trim = 0mm 0mm 0mm 0mm, clip, scale=0.5]{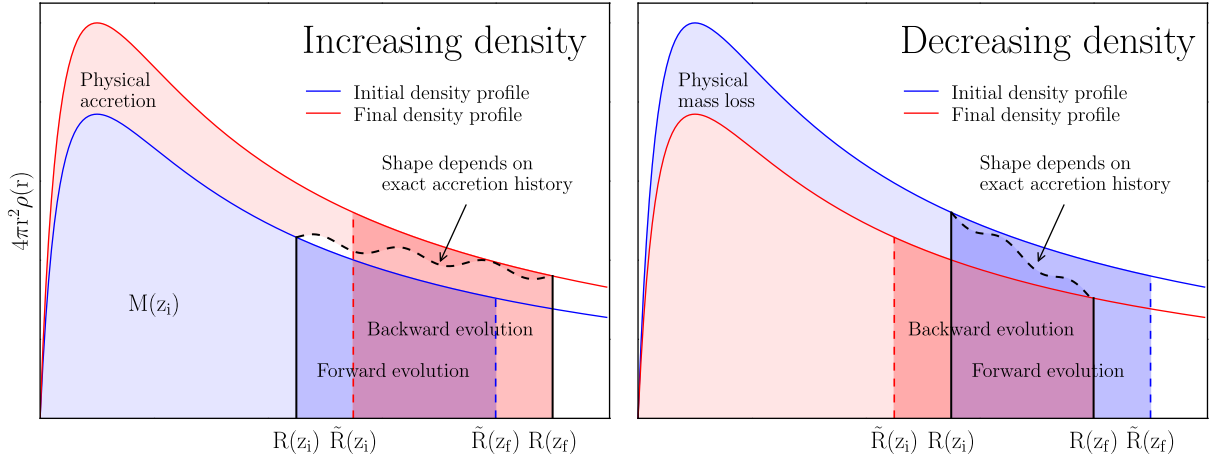} 
\caption{Visualization of pseudo-evolution in the presence of physical accretion (left panel) and mass loss (right panel). Left panel: the blue and red lines correspond to an NFW density profile which has uniformly increased from redshift $z_\rmi$ (blue) to redshift $z_\rmf$ (red). The original halo mass corresponds to the area under the blue curve and inside $R(z_\rmi)$. The true pseudo-evolution corresponds to the area between the thick, black lines: the true virial radii at $z_\rmi$ and $z_\rmf$, as well as a line which depends on which parts of the halo fell in, and which were static when the virial radii crossed their position. Forward evolution corresponds to the darker blue shaded area under the blue curve, and between $R(z_\rmi)$ and $\tilde{R}(z_\rmf)$, the virial radius at $z_\rmf$ as estimated using the profile at $z_\rmi$. Because both $\tilde{R}(z_\rmf)$ and the density profile are underestimated, the amount of pseudo-evolution is underestimated. For backward evolution (dark red shaded area), there are two competing effects which mean it might under- or overestimate the true amount of pseudo-evolution. Right panel: the case of mass loss is less frequent than mass growth, but not irrelevant. In this scenario, the virial radius at $z_\rmf$ is still larger than at $z_\rmi$, despite the mass loss. Due to pseudo-evolution, this is always true in practice. In this case, forward evolution invariably {\it overestimates} the true amount of pseudo-evolution, while backward evolution may overestimate or underestimate it. See Section \ref{subsec:nonstatic} for further discussion.}
\label{fig:notstatic}
\end{figure*}

In the last section, we considered the effects of pseudo-evolution in the simple case where a halo profile does not undergo any physical evolution. As we show below, such halos are indeed abundant at low redshifts, but there are of course also many halos which do undergo actual physical evolution. Interpreting the mass evolution of such halos, and estimating the contribution from pseudo-evolution, is somewhat trickier. In this section, we use simple toy models for the evolution of density profiles, and investigate the contribution from pseudo-evolution in such cases.  

Over a small redshift range, the mass evolution can be split into two terms, 
\begin{equation}
\label{eq:mdeltaderiv1}
\frac{\rmd M_\Delta}{\rmd z} = 4\pi R_\Delta^2\,\rho(R_\Delta,z)\,\frac{\rmd R_\Delta}{\rmd z} + \int_{0}^{R_\Delta(z)} 4 \pi r^2\, \frac{\rmd \rho(r)}{\rmd z} \rmd r\,,
\end{equation}
where the first term corresponds to the pseudo-evolution at redshift $z$ due to the changing virial radius, and the second term represents the actual physical growth of the halo due to accretion. The total evolution of the halo mass is then the integral of the above equation from an initial redshift $z_\rmi$ to a final redshift $z_\rmf$, where $z_\rmi>z_\rmf$. For clarity, we drop the $\Delta$ subscripts, and it is understood that $M$ stands for the virial mass $M_{\Delta}$ and $R$ for the virial radius $R_{\Delta}$. Integrating the differential Equation (\ref{eq:mdeltaderiv1}), we get
\begin{equation}
M(z_\rmf) = M(z_\rmi) + \Delta M_{\rm pseudo}(z_\rmf) + \Delta M_{\rm phys}(z_\rmf) \,,
\end{equation}
where
\begin{eqnarray}
\label{eq:def_dmpseudo}
\Delta M_{\rm pseudo}(z_\rmf) &=& \int_{R(z_\rmi)}^{R(z_\rmf)} \rmd r \,4 \pi r^2 \,\rho(r, z_\rmc) \,,\\
\label{eq:def_dmphys}
\Delta M_{\rm phys}(z_\rmf) &=& \int_{z_\rmi}^{z_\rmf} \rmd z \int_{0}^{R(z)} \rmd r\, 4 \pi r^2 \frac{\rmd \rho(r, z)}{\rmd z} \,.
\end{eqnarray}
Note that we have retained the redshift dependence of the virial radius and density in the first equation to make it explicit that the density at position $r$ should correspond to the epoch when the boundary crosses $r$. For clarity, we denote the redshift at crossing $z_\rmc$.

When evaluating pseudo-evolution for halos identified in simulations, we have to resort to using the density profiles of halos at fixed instants (snapshots), which may be widely spaced in time. For truly static density profiles, this poses no problem, because we can evaluate Equation (\ref{eq:def_dmpseudo}) using the profile at either the initial or the final snapshot. If the density profile is not static, however, Equations (\ref{eq:def_dmpseudo}) and (\ref{eq:def_dmphys}) imply that a particle is added to the physically accreted mass if it fell into the virial radius between $z_\rmi$ and $z_\rmf$, and to the pseudo-evolution if it was stationary when the virial radius crossed its position in space. Such a proper estimate of the amount of pseudo-evolution is possible, provided simulation snapshots sufficiently finely spaced in time are available. However, in practice this is not always the case and one has to work with a limited number of snapshots. 

To estimate the amount of pseudo-evolution between any two snapshots at $z_\rmi$ and $z_\rmf$, let us first use the density profile at the initial redshift $z_\rmi$. We call this estimate {\it forward evolution}. For ease of interpretation, we split the pseudo evolution integral as follows,
\begin{eqnarray}
\label{eq:def_forward}
\Delta M_{\rm pseudo} (z_\rmf) &=& \int_{R(z_\rmi)}^{\tilde{R}(z_\rmf)} \rmd r \,4 \pi r^2 \,\rho(r, z_\rmi) \nonumber \\ 
                         &&+ \int_{R(z_\rmi)}^{\tilde{R}(z_\rmf)} \rmd r \,4 \pi r^2 \,[\rho(r, z_\rmc)-\rho(r, z_\rmi)] \nonumber \\
                         &&+ \int_{\tilde{R}(z_\rmf)}^{R(z_\rmf)} \rmd r \,4 \pi r^2 \,\rho(r, z_\rmc) \,.
\end{eqnarray}
Here the first integral represents the forward evolution estimate using the density profile at redshift $z_\rmi$, while the other two terms cannot be calculated without knowledge of the density profile at intermediate redshifts $z_\rmc$. Note that we integrate the first term only out to the radius $\tilde{R}(z_\rmf)$, which denotes the boundary of the halo at $z_\rmf$ inferred based upon the density profile at redshift $z_\rmi$. If the density profile is truly static the other two terms vanish. If the density profile increases (this is the more common case, but see Section \ref{subsec:individual} for exceptions), the latter two terms are positive ($\rho(r, z_\rmc)>\rho(r, z_\rmi)$ and $\tilde{R}(z_\rmf)<R(z_\rmf)$), and the first term thus {\it underestimates} the amount of pseudo-evolution. This scenario is visualized in the left panel of Figure \ref{fig:notstatic}, with the forward evolution estimate shown as the darker blue shaded area. If the density profile, however, for some reason {\it decreases}, forward evolution always {\it overestimates} the true amount of pseudo-evolution, as shown in the right panel of Figure \ref{fig:notstatic}.

Alternatively, we can use the density profile at the final redshift $z_\rmf$ to predict the amount of pseudo-evolution that occurred between $z_\rmi$ and $z_\rmf$. We call this estimate {\it backward evolution}. In this case, the integral can be split as
\begin{eqnarray}
\label{eq:def_backward}
\Delta M_{\rm pseudo}(z_\rmi) &=& \int_{\tilde{R}(z_\rmi)}^{R(z_\rmf)} \rmd r \,4 \pi r^2 \,\rho(r, z_\rmf) \nonumber \\ 
                         &&+ \int_{\tilde{R}(z_\rmi)}^{R(z_\rmf)} \rmd r \,4 \pi r^2 \,[\rho(r, z_\rmc)-\rho(R, z_\rmf)] \nonumber \\
                         &&+ \int_{R(z_\rmi)}^{\tilde{R}(z_\rmi)} \rmd r \,4 \pi r^2 \,\rho(r, z_\rmc) \,,
\end{eqnarray}
where $\tilde{R}(z_\rmi)$ is the boundary at $z_\rmi$ as inferred by using the density profile at $z_\rmf$. Again, the first term represents the backward evolution estimate, and the latter two terms vanish if the density profile is truly static. If the profile grows, the second term is negative and the third term is positive as $R(z_\rmi)<\tilde{R}(z_\rmi)$. Thus, we cannot be certain whether the backward evolution underestimates or overestimates the true amount of pseudo-evolution. This is visualized in the left panel of Figure \ref{fig:notstatic}, where the backward evolution estimate is shown as the darker red shaded area. While backward evolution misses part of the true pseudo-evolution ($R(z_\rmi)<r<\tilde{R}(z_\rmi)$), it also includes a component that should be attributed to physical accretion (the area above the dotted line). The same is true if the profile physically loses mass. In that case, backward evolution underestimates the density profile ($\rho(r,z_c)>\rho(r,z_\rmf)$), but at the same time also underestimates the virial radius at $z_\rmi$, as shown in the right panel of Figure \ref{fig:notstatic}. Therefore, backward evolution can end up being either smaller or larger than the true amount of pseudo-evolution, depending on which of the two competing effects dominates.

For the forward and backward estimates of pseudo-evolution, we only used one density profile for each case. This leads to an extra error term, because we under- or overestimate the true virial radius at $z_\rmf$ or $z_\rmi$, depending on whether we use forward or backward evolution. From Figure \ref{fig:notstatic}, it is clear that, given the density profiles at both $z_\rmi$ and $z_\rmf$, we can improve our estimate by using the {\it true} virial radii, $R(z_\rmi)$ and $R(z_\rmf)$, instead of the estimated ones, $\tilde{R}(z_\rmi)$ and $\tilde{R}(z_\rmf)$. The true amount of pseudo-evolution is represented by the area inside the thick, black lines in Figure \ref{fig:notstatic} (the true virial radii, and the dashed line which depends on the halo's accretion history). Moreover, as long as the profile grows {\it monotonically} and {\it at all radii}, the dashed black line in Figure \ref{fig:notstatic} lies between the profiles at $z_i$ and $z_f$ (as depicted), and we can write down definite lower and upper limits for the amount of pseudo-evolution,
\begin{eqnarray}
\label{eq:minmax}
\Delta M_{\rm min}(z_\rmi,z_\rmf) &=& \int_{R(z_\rmi)}^{R(z_\rmf)} \rmd r \,4 \pi r^2 \,\rho(r, z_\rmi) \nonumber \\
\Delta M_{\rm max}(z_\rmi,z_\rmf) &=& \int_{R(z_\rmi)}^{R(z_\rmf)} \rmd r \,4 \pi r^2 \,\rho(r, z_\rmf) \,,
\end{eqnarray}
where the lower and upper limits are exchanged if the profile is decreasing at all radii. The difference between the upper and lower limits corresponds to the area between the red and blue curves, and the true virial radii $R(z_\rmi)$ and $R(z_\rmf)$ (black lines) in Figure \ref{fig:notstatic}. Of course, depending on the amount of physical growth, this area can be large, and the shape of the dashed black line is, in principle, not known without knowledge of the density profile at all intermediate redshifts.

However, for the case of modest physical growth, we can estimate the shape of this line. Let us assume that the density follows an NFW profile at all times, that the profile grows by the same fractional amount at all radii, and that this growth is linear with cosmic time $t(z)$, corresponding to a simple re-scaling of the NFW profile as
\begin{equation}
\rho_s(z) = \rho_s(z_\rmi) \left(1+ \frac{t_\rmf - t(z)}{t_\rmf - t_\rmi} g_\rmf \right) \,,
\end{equation}
where $g_\rmf$ is the final amount of growth compared to $z_\rmi$. Such a model seems justified from the mean density evolution observed in simulations (see Figure \ref{fig:directcomp}). Furthermore, we assume that the scale radius stays constant (see e.g. \citet{bullock_01_halo_profiles} for justification within the relevant redshift range, as discussed in Section\ref{subsec:zrange}). We can now compute the fraction of the difference between $\Delta M_{\rm min}$ and $\Delta M_{\rm max}$ which should be counted as pseudo-evolution,
\begin{equation}
f = \frac{\int_{R(z_\rmi)}^{R(z_\rmf)} \rmd r \,4 \pi r^2 \,[\rho(r, z_\rmc)-\rho(r, z_\rmi)]}{\int_{R(z_\rmi)}^{R(z_\rmf)} \rmd r \,4 \pi r^2 \,[\rho(r, z_\rmf)-\rho(r, z_\rmi)]}\,.
\end{equation}
Equation (\ref{eq:mu}) can be used to evaluate $f$ numerically as an integral over redshift,
\begin{equation}
f = \frac{\int_{z_\rmi}^{z_\rmf} 4 \pi R(z)^2 [\rho(R[z], z)-\rho(R[z], z_\rmi)] \frac{\rmd R}{\rmd z} \rmd z}{\int_{R(z_\rmi)}^{R(z_\rmf)} \rmd r \,4 \pi r^2 \,[\rho(r, z_\rmf)-\rho(r, z_\rmi)]}\,.
\end{equation}
Besides the initial and final redshifts, we need to specify an initial halo mass, $M(z_\rmi)$,  an initial concentration, $c_\rmi$, and the growth factor, $g_\rmf$.  For a wide range of reasonable values of these parameters, $f$ is approximately constant  regardless of whether halo profile increases or decreases (corresponding to positive and negative values of $g_\rmf$). In order to be conservative and avoid overestimating the amount of pseudo-evolution, we use the lowest values of $f$ found for any combination of $M(z_\rmi)$, $c_\rmi$ and $g_\rmf$. Furthermore, we investigated linear growth with $z$ instead of $t(z)$, and find that it consistently leads to larger values of $f$. Again, we choose the lower values of $f$ from linear growth in $t(z)$. We do, however, find that $f$ depends on the chosen mass definition. Using the aforementioned lowest values, we find $f_{200\bar{\rho}} \approx 0.45$, $f_{\rm vir} \approx 0.4$, and $f_{500\rho_\rmc} \approx 0.34$. We can now write down our best estimator of the true amount of pseudo-evolution,
\begin{equation}
\label{eq:def_best}
\Delta M_{\rm best}(z_\rmf) = (1-f)\, \Delta M_{\rm min}(z_\rmf) + f\, \Delta M_{\rm max}(z_\rmf)\,.
\end{equation}
Besides making some strong assumptions as to the mode of halo growth, this estimator does {\it not} capture the effects of mergers which entirely contribute to physical accretion. Thus, one should always refer to $\Delta M_{\rm min}$ as a safe lower estimate of pseudo-evolution. However, for the case of gradual, uniform halo growth, $\Delta M_{\rm best}$ should be a reasonable approximation.

We have now derived five different estimators of pseudo-evolution, two of which use one density profile only (forward and backward, Equations (\ref{eq:def_forward}) and (\ref{eq:def_backward})), as well as three estimators which use two density profiles (minimum, maximum and the best estimator, Equations (\ref{eq:minmax}) and (\ref{eq:def_best})). We focus on the forward and backward estimates in Section \ref{subsec:mah200b}, and investigate the difference between the estimators quantitatively in Section \ref{subsec:individual}.

\subsection{The Relevant Redshift Range}
\label{subsec:zrange}

From the discussion in Section \ref{subsec:staticmodel} it is clear that pseudo-evolution {\it always} occurs, as long as the reference density $\bar{\rho}$ changes. However, while a physical halo density profile evolves rapidly, \citep[{\it fast accretion mode},][]{zhao_03_mah}, it is difficult to disentangle the effects of pseudo-evolution and physical accretion (see Section \ref{subsec:nonstatic}). Thus, we focus on a redshift range, and halo mass range, where halos are mostly in the {\it slow accretion mode}.

The pace of accretion is a function of the ratio of $f=M/M_{NL}$ (or more generally the peak height $\nu$), where $M_{NL}$ is the characteristic mass scale of fluctuations that undergo collapse at redshift $z$ \citep{kravtsov_12_cluster_review}. $M_{NL}$ depends on the linear growth rate, which in turn depends primarily on $\Omega_\rmm$, and to a lesser extent on $\Omega_{\Lambda}$ \citep[see e.g. Figure 1 in][]{hamilton_01_growthfactor}. For $z < 1$, galaxy-sized halos ($M \approx 10^{12}\ \rm M_{\odot}$) enter the slow accretion regime, whereas cluster-sized halos ($M>10^{14}\ \rm M_{\odot}$) are mostly still in the fast accretion mode today.

Thus, for the purposes of this paper, we focus on the mass evolution from $z=1$ to $z=0$. We emphasize that this does not mean that pseudo-evolution does not contribute to the halo mass growth at higher redshifts. 

%%%%%%%%%%%%%%%%%%%%%%%%%%%%%%%%%%%%%%%%%%%%%%%%%%%%%%%%%%%%%%%%%%%%%%%%%%
% MASS EVO FROM SIMULATIONS
%%%%%%%%%%%%%%%%%%%%%%%%%%%%%%%%%%%%%%%%%%%%%%%%%%%%%%%%%%%%%%%%%%%%%%%%%%

\section{Halo mass evolution in simulations}
\label{sec:profiles}

\subsection{Numerical Simulation}
\label{subsec:sim}

To quantify the pseudo-evolution of mass using realistic halo profiles, we use a sample of halos extracted from a dissipationless cosmological simulation of the $\Lambda$CDM model. Specifically, we use  the Bolshoi simulation \citep{klypin_11_bolshoi}, which followed the evolution of the matter distribution using the Adaptive Refinement Tree (ART) code \citep{kravtsov_97_art, gottloeber_08_art} in a flat $\Lambda$CDM model with parameters $\Omega_\rmm=1-\Omega_\Lambda=0.27$, $\Omega_\rmb=0.0469$, $h=H_0/(100\kmsmpc)=0.7$, $\sigma_8=0.82$ and $n_\rms=0.95$. These cosmological parameters are compatible with measurements from WMAP7 \citep{jarosik_11_wmap7}, a combination of WMAP5, Baryon Acoustic Oscillations, and supernovae \citep[SNe;][]{komatsu_etal11}, X-ray cluster studies \citep{vikhlinin_09_clusters}, and observations of the clustering of galaxies and
galaxy-galaxy/cluster weak lensing \citep[see e.g.][]{tinker_12_galcosmo, more_12_galaxy_dm}. The same cosmology was used for the calculations shown in Figure \ref{fig:analytical_prediction}, and will be used for the remainder of this paper.

The Bolshoi simulation uses $2048^3\approx 8$ billion particles to follow the evolution of the matter distribution in a cubic box of size $250\mpch$, which corresponds to a particle mass of $1.35\times 10^8 \msunh$. This implies that the smallest halos considered in this paper ($\mvir = 2\times 10^{11} \msunh$) are resolved by over 1000 particles. As density peaks, we use the centers of halos from a catalog generated with the bound density maxima (BDM) algorithm \citep{klypin_97_bdm, klypin_11_bolshoi}. 

We identified all distinct halos with $\mvir \geq 2 \times 10^{11} \msun$ from the simulation at $z = 0$, resulting in a sample of about $240,000$ halos. For each of these halos, we constructed radial density profiles by summing the particle contributions in 80 logarithmically spaced bins, spanning radii from $0.05$ to $10\ \rvir$. We have checked that our profiles are in excellent agreement with the density profiles for the same Bolshoi halos extracted from the MultiDARK database \citep{riebe_11_md_database}. The larger radial range and finer resolution of our profiles compared to the ones existing on the database allowed us to define halo masses for lower overdensity thresholds of $\Delta_{\rmm} = 200$, and much lower background densities. 

For the analysis of individual halos in Section \ref{subsec:individual}, we needed to match individual isolated halos to their progenitors at $z=1$. To identify progenitors of $z=0$ halos at $z=1$, we rank order particles in each halo by their binding energy and consider $20\%$ of the most bound particles. A $z=1$ halo is deemed a progenitor of a $z=0$ halo if it has more than half of its most bound particles among the most bound particles of the $z=0$ halo. The binding energy of particles within $\rvir(z=0)$ was estimated as
\begin{equation}
E_b =  \frac{1}{2}(v_x^2+v_y^2+v_z^2)+\Phi_{NFW}(r)
\end{equation}
where $v_x$, $v_y$, $v_z$ are the components of the particle velocity from the simulation output. The potential was estimated assuming an NFW profile,
\begin{eqnarray}
\Phi_{NFW}(r) &=& -4 \pi G \rho_s r_s^2 \frac{\ln{(1+r/r_s)}}{r/r_s} \nonumber \\
&=&-4.625 v_{max}^2 \frac{\ln{(1+r/r_s)}}{r/r_s}
\end{eqnarray} 
where $v_{max}$ is the maximum circular velocity and $r_s$ calculated as $r_s=r_{\rm max}/2.16$, where $r_{\rm max}$ is the radius at which $v_{\rm max}$ is reached. Both $v_{\rm max}$ and $r_{\rm max}$ are provided for each halo in the BDM catalog.  For a few percent of halos, this method fails to identify a progenitor, indicating that the halo was assembled from many different, generally much smaller halos. We discard such halos, because it would be very difficult to get a meaningful estimate of pseudo-evolution for such halos anyway. We identified progenitors for a small subset of the halos in the Bolshoi simulation, namely those in the three narrow mass bins discussed in Sections \ref{subsec:individual} and \ref{subsec:nonevo}.

\begin{figure*}
\centering
\includegraphics[trim = 0mm 0mm 58mm 0mm, clip, scale=0.6]{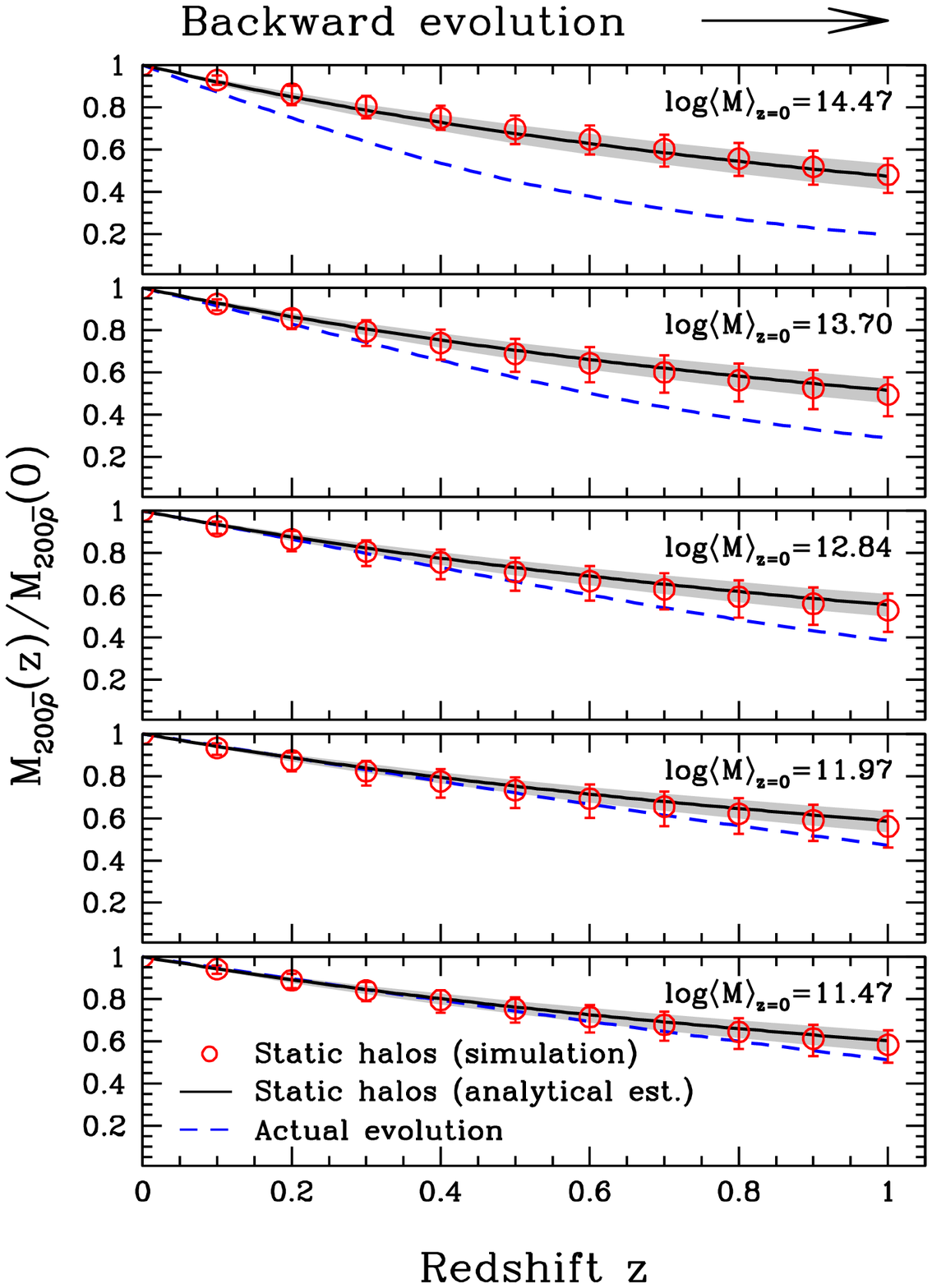} 
\includegraphics[trim = 0mm 0mm 58mm 0mm, clip, scale=0.6]{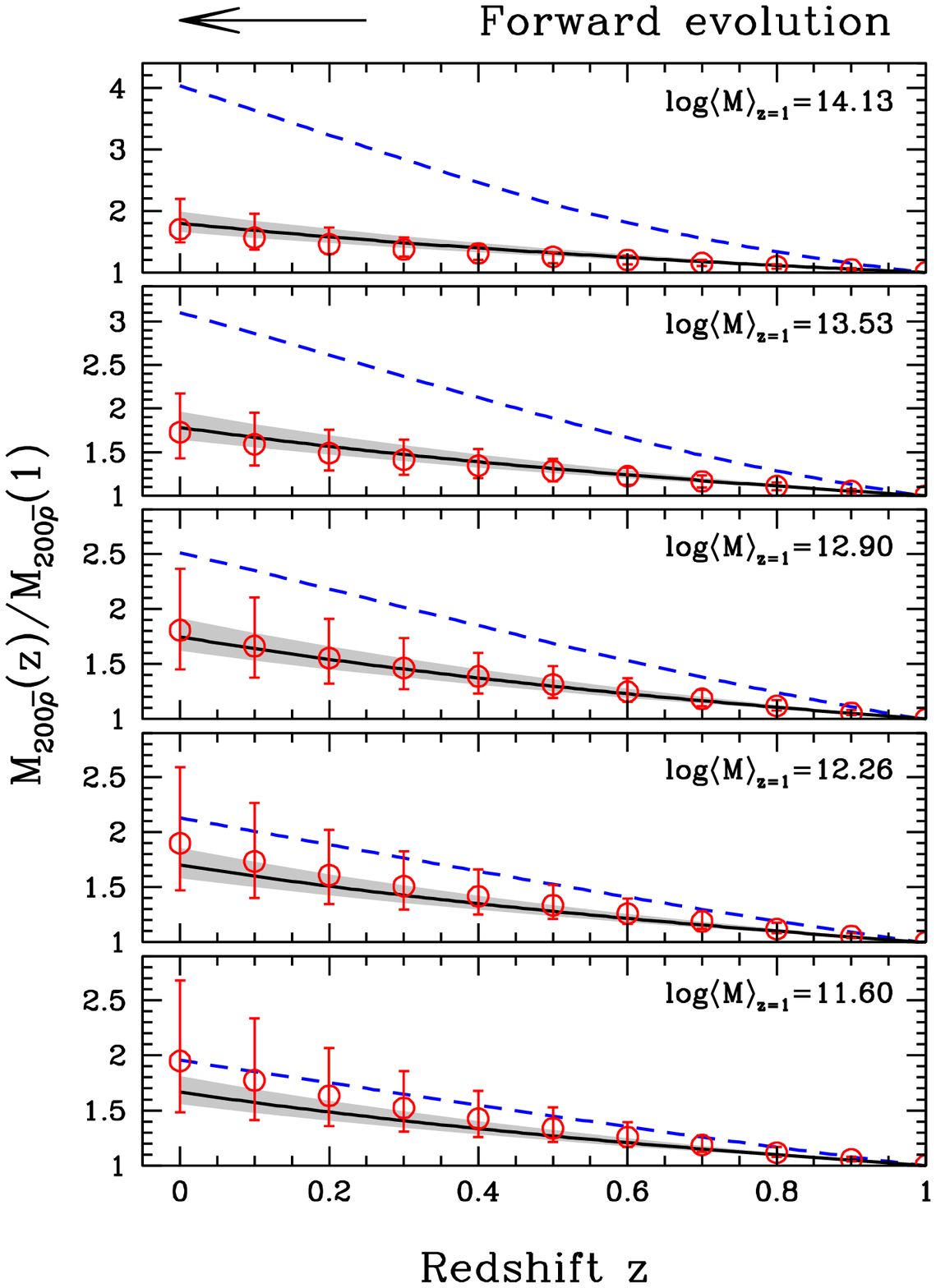} 
\caption{Mass evolution for different halo masses for the $\mmean$ mass definition, in five logarithmic mass bins. The dashed lines show the actual mass evolution of halos as predicted by the Z09 model. The solid lines show the evolution of mass due solely to the evolution of the reference density in the mass definition ({\it pseudo-evolution}) as predicted by the static halo evolution model (see Section \ref{subsec:staticmodel}). The gray band around the solid lines shows the 68\% confidence interval due to scatter in the concentration-mass relation. The red points show the  pseudo-evolution computed using density profiles from the Bolshoi simulation, with error bars indicating the mass range containing 68\% of the halos in a given mass bin. The left panels show this evolution computed by extrapolating the mass evolution using profiles at $z=0$ and going backwards in time, while the right panels show the evolution of profiles at $z=1$, going forward in time (see Section \ref{subsec:nonstatic}). Note, however, that the mass bins on the left and right do not correspond to the same halo masses, and should thus not be compared directly. The scale for the top two panels on the right differs from the bottom three panels. These results demonstrate that pseudo-evolution accounts for at least half of mass evolution at all halo masses. For small halos ($\mmean \simlt 10^{12} \msunh$), most of the mass change from $z=1$ to $z=0$ is due to pseudo-evolution. See Section \ref{subsec:mah200b} for further discussion.}
\label{fig:mah200b}
\end{figure*}

\begin{figure*}
\centering
\includegraphics[trim = 0mm 0mm 58mm 105mm, clip, scale=0.6]{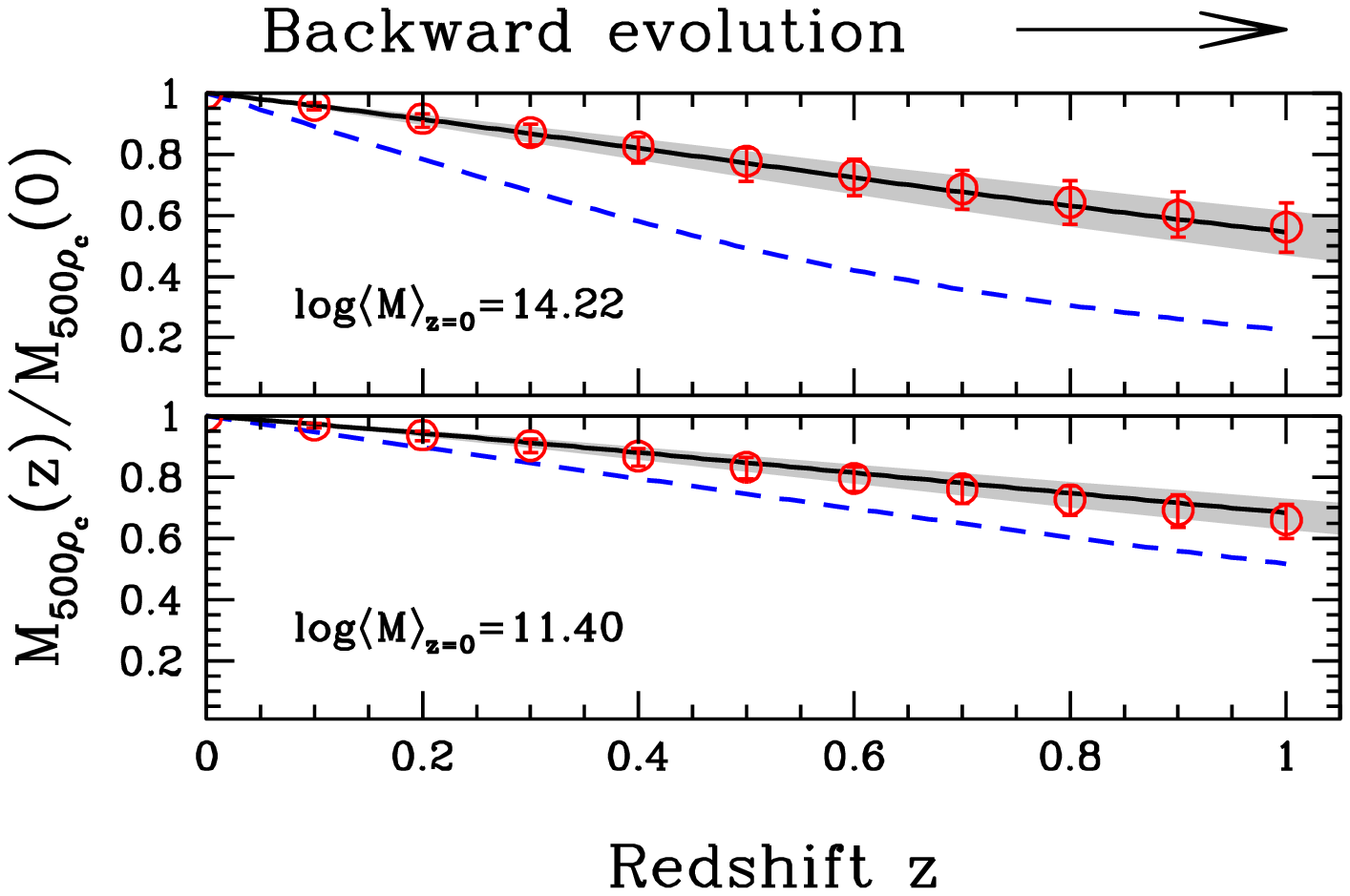} 
\includegraphics[trim = 0mm 0mm 58mm 105mm, clip, scale=0.6]{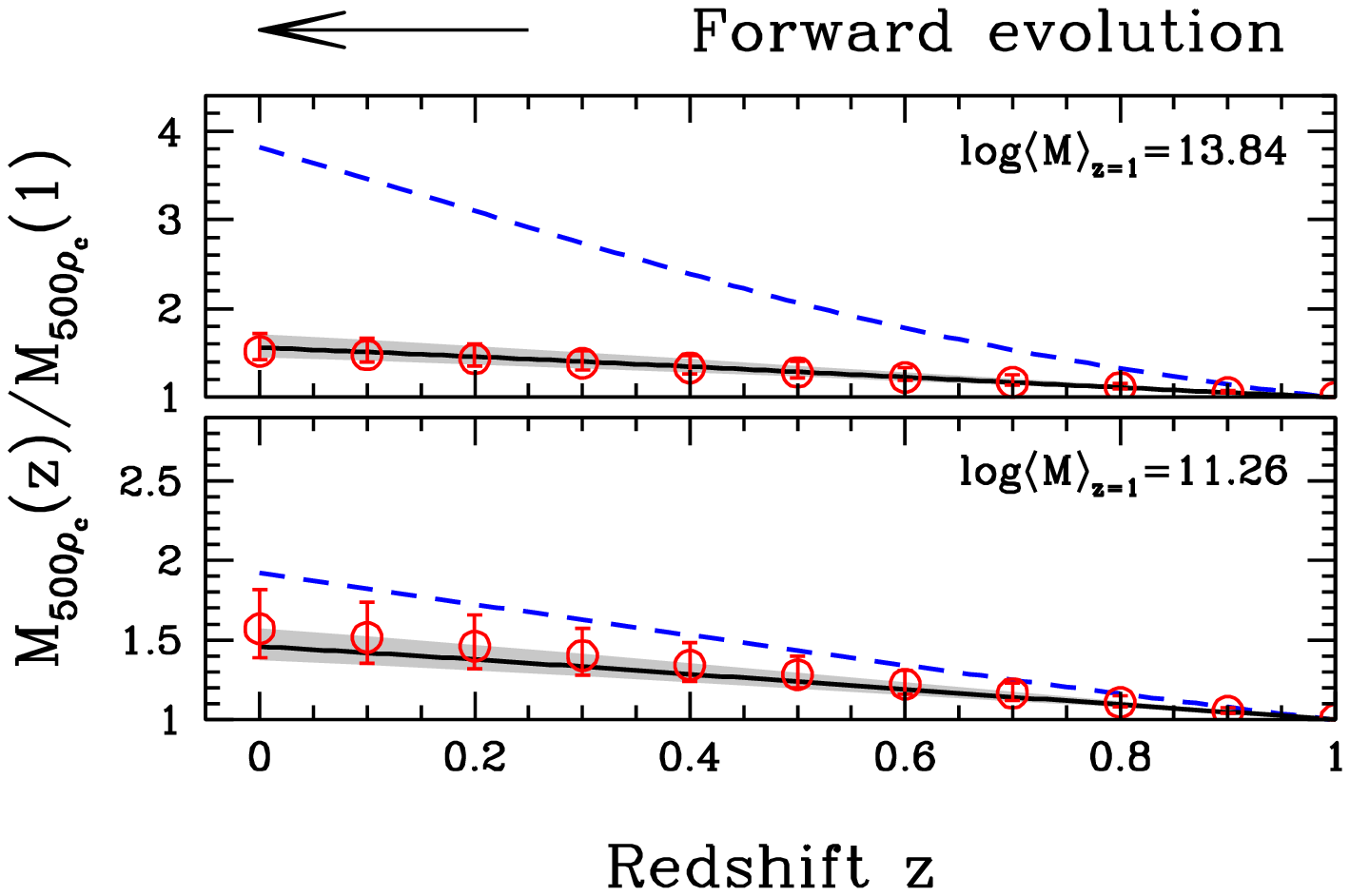} 
\caption{Same as Figure \ref{fig:mah200b}, but for the $\mcrit$ mass definition. Only the highest and lowest of five mass bins are shown. The analytical prediction of the static halo model and the results from simulated halos match even better than for $\mmean$, and the Bolshoi results exhibit smaller scatter. This is caused by a smaller virial radius $\rcrit$, meaning that irregularities in the outskirts of halos play a lesser role. The pseudo-evolution is a little weaker in $\mcrit$ than in $\mmean$, but still accounts for most of the mass evolution at low halo masses.}
\label{fig:mah500c}
\end{figure*}

\subsection{The Mean Evolution of $\mmean$ and $\mcrit$}
\label{subsec:mah200b}

We first focus on mass definitions based on the mean matter density of the universe, such as  $\mmean$. To estimate the pseudo-evolution of halo mass using realistic matter density profiles, we assume that the density profiles stay constant in physical units from $z=1$ to $z=0$ and evolve the background density according to 
\begin{equation}
\label{eq:bgdensity}
\bar{\rho}(z) = (1+z)^3 \bar{\rho}(0)\,.
\end{equation}
The radius, $\rmean$, of the halo is then numerically identified to be the radius which encloses an average overdensity of $\Delta=200$ with respect to $\bar{\rho}(z)$, and the mass, $\mmean$, follows from Equation (\ref{eq:mdelta1}). 
Following the discussion in Section \ref{subsec:nonstatic}, we consider both backward and forward evolution by predicting the amount of pseudo-evolution from the density profiles at $z=0$ and $z=1$, respectively. We emphasize that we do {\it not} use merger trees for this simple estimate, because we are only trying to quantify the mean pseudo-evolution in $\mmean$, not the evolution of individual halos which we will tackle in Section \ref{subsec:individual}.

In Figure \ref{fig:mah200b}, we show the results of backward evolution (left panels) and forward evolution (right panels). The mean of the ratio of the evolved halo mass to the mass at the profile redshift is plotted using open circles, and the error bars indicate the 16 and 84 percentiles of this distribution. The analytical estimate from Section \ref{subsec:staticmodel} is shown using a solid line, with gray contours indicating the 1$\sigma$ scatter in the analytical prediction. This scatter arises due to scatter in the concentration-mass relation, which we assumed to be 0.14 dex based on the results of \citet{wechsler_02_halo_assembly}. The excellent agreement between the analytical estimate and the results from the halo profiles implies that our assumption about the density distribution at redshift zero (from the models of Z09) is not too far off from the actual density distribution of halos in the simulation. This shows that the analytical model can provide an excellent description of the mass evolution if the physical density distribution of the halos was indeed constant. Furthermore, the good agreement with the static halo model shows that those halos do indeed follow the NFW density profiles. The agreement is better for backward evolution than for forward evolution, indicating that deviations from the NFW form of the density profiles of halos are larger at $z=1$ \citep[e.g.,][]{tasitsiomi_04_clusterprof}.

Next, we would like to contrast the predictions of forward and backward evolution with the {\it true} mass evolution histories of halos observed in simulations. We make use of the Z09 model for the mass evolution histories of halos which has been shown to accurately reproduce the mass evolution histories for a large variety of cosmological models (scale-free or $\Lambda$CDM). The results from this model will thus include the effects of both pseudo-evolution and the actual physical accretion of mass. By comparing our estimates of pseudo-evolution to these realistic mass evolution histories, we can disentangle the two effects. The mass evolution histories predicted by the model of Z09 are shown by dashed lines in Figure \ref{fig:mah200b}. 

As expected from the discussion in Section \ref{subsec:nonstatic}, backward evolution predicts a larger amount of pseudo-evolution than forward evolution at all halo masses, and this difference increases with halo mass (because larger halos undergo more physical accretion). For low-mass halos ($\mmean \lsim 10^{12}\msunh$), backward and forward evolution agree, as can be expected for completely non-evolving density profiles. Both predict that pseudo-evolution accounts for almost all of the mass change of low-mass halos since $z=1$. Because forward evolution can only underpredict the true amount of pseudo-evolution, this result clearly shows that the density profiles of low-mass halos are on average already established at $z=1$ and change very little with time. 

For cluster-sized halos ($\mmean \sim 10^{14}\msunh$), backward evolution predicts about $60\%$ of the mass evolution to be pseudo-evolution, but forward evolution predicts that about a third of mass change since $z=1$ is due to pseudo-evolution on average. As we noted above, such differences are expected for profiles that do physically evolve, as cluster halos do. In this case, it is difficult to deduce from the MAHs in Figure \ref{fig:mah200b} how much of their evolution is actually due to pseudo-evolution. However, the forward and backward evolution estimates only used one snapshot each. In order to quantify pseudo-evolution more carefully and to shed light on the pseudo-evolution of large halos, we investigate forward and backward evolution for individual halos, as well as the more advanced estimators discussed in Sections \ref{subsec:nonstatic} and \ref{subsec:individual}.

The mass evolution shown in Figure \ref{fig:mah200b} refers to a sample of isolated halos only. Inevitably, this sample contains close pairs of halos which are not identified as overlapping in the halo catalog, but whose density profiles pick up contributions from particles belonging to the other halo. As the density profiles we use extend as far as $10\ \rvir$, this is the case for a significant number of halos in our sample. When we evolve such a halo forward in time, its radius grows, and the contribution from other halos leads to excess mass growth, which manifests itself as a large scatter in the mass evolution history. However, many of these halos end up lying inside the radius of a neighboring, larger halo, and we need to exclude them from the averaged mass evolutions shown in Figure \ref{fig:mah200b}. 

In principle, the most straightforward way to identify subhalos would be to use merger trees of the Bolshoi simulation. However, merger trees are generated with knowledge of the full mass evolution of halos (rather than just the pseudo-evolution), as well as their motion and acceleration. In the spirit of our extremely simple model of static halo profiles, we wish to avoid using such information, and rely only on the density profiles and initial positions of the halos in our sample. 

In the case of backward evolution, halo radii decrease, meaning that we do not need to worry about halos becoming subhalos. In the case of forward evolution, we identify subhalos as follows. At each redshift, we evolve the virial radii of all halos to match the evolved reference density. We then check whether the new virial radius of such a halo encloses the center of any other, smaller halos. Note that we assume that the neighboring halos stay at a constant physical distance and are not part of the Hubble flow, consistent with our assumption that the physical density around the peak does not evolve. We exclude the subhalos discovered in this manner from the current redshift bin, and all subsequent smaller redshifts. We start this process with the largest halo in the sample, marching down to the smallest halos. Once a halo has been found to be a subhalo, it cannot itself be the host of another halo. By the end of the evolution from $z=1$ to $z=0$, a total of about 14\% of the halos in the sample had become subhalos and been removed from the sample. The mass evolution averages shown in Figure \ref{fig:mah200b} are insensitive to the removal of subhalos, but the scatter is reduced significantly by this procedure. 

Following our discussion of mass definitions based on the mean matter density of the universe such as $\mmean$, we now investigate definitions based on the critical density. Note that the difference between those definitions is not only due to the different values for $\Delta$ which are typically chosen, but that $\rho_\rmc$ evolves qualitatively different from $\bar{\rho}$ such that
\begin{equation}
\rho_\rmc(z)=\bar{\rho}(z) \left[ \frac{E^2(z)}{\Omega_{m,0} (1+z)^3} \right].
\end{equation}
Figure \ref{fig:mah500c} shows the evolution histories of $\mcrit$ for two mass bins, compared to the true evolution represented by the model presented by Z09. The results were derived in exactly the same way as the results for $\mmean$, except for the different evolution of $\Delta \rho_{\rm ref}$. The pseudo-evolution is slightly weaker in $\mcrit$ than $\mmean$. This can be seen by comparing, for example, the lowest mass bins in Figs.~\ref{fig:mah200b} and \ref{fig:mah500c}. The weaker evolution in $\mcrit$ may seem slightly counter-intuitive at first, because $\rcrit$ is {\it smaller} than $\rmean$ (which implies that $c_{\rm 500\rho_\rmc}< c_{200\bar\rho} $), and mass profiles as a function of enclosed density tend to steepen towards the center of halos (see Figure \ref{fig:static}). However, the weaker evolution of $\rho_\rmc$ compared to $\bar{\rho}$ more than offsets this effect. For example, at $z=1$ the mean matter density of the universe was a factor of eight higher than today, but the critical density was only larger by a factor of $E^2(1) \approx 2.9$. 

\subsection{The pseudo-evolution of individual halos}
\label{subsec:individual}

\begin{figure*}
\centering
\includegraphics[trim = 0mm 0mm 8mm 10mm, clip, scale=0.47]{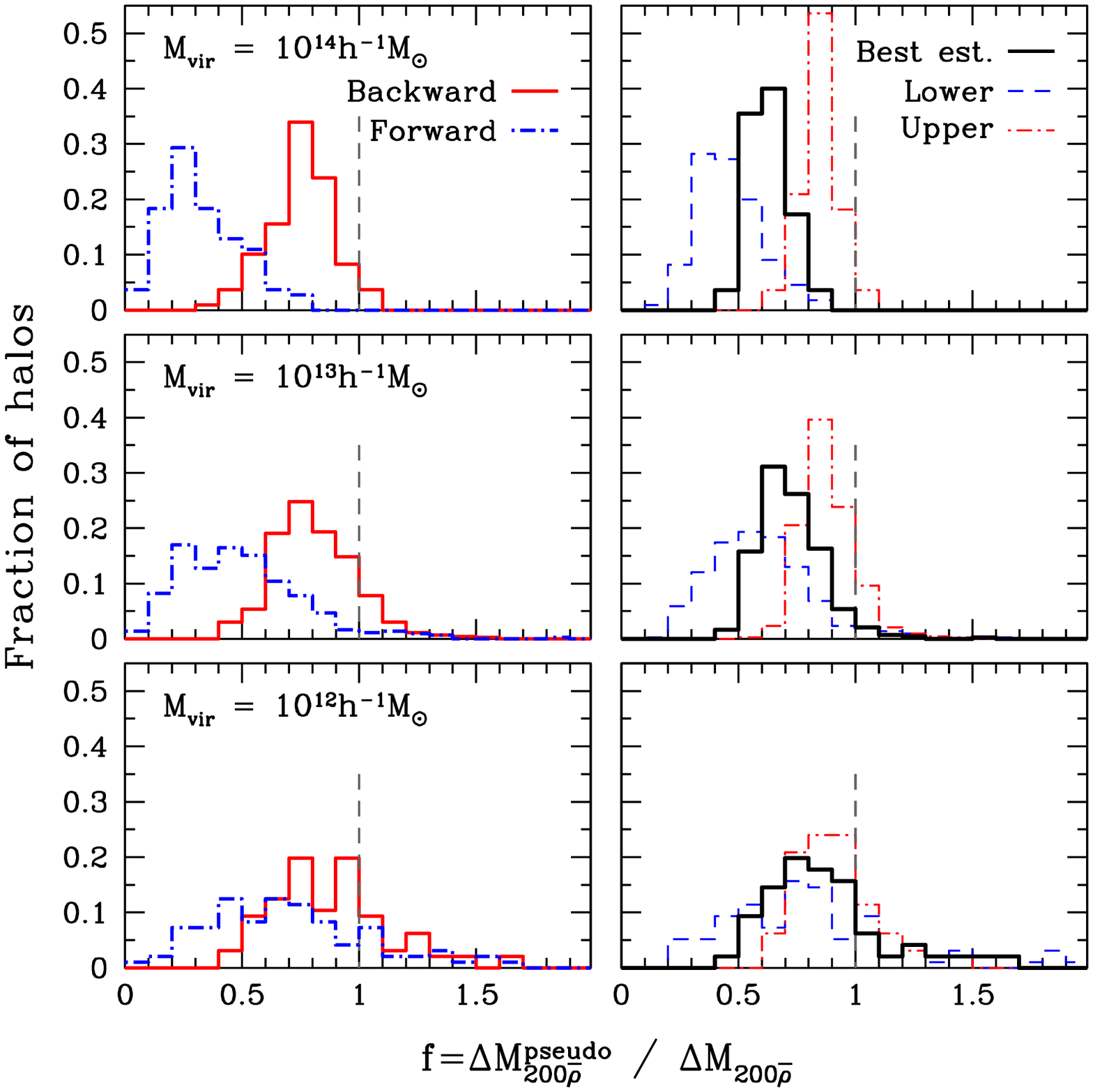} 
\includegraphics[trim = 20mm 0mm 5mm 10mm, clip, scale=0.47]{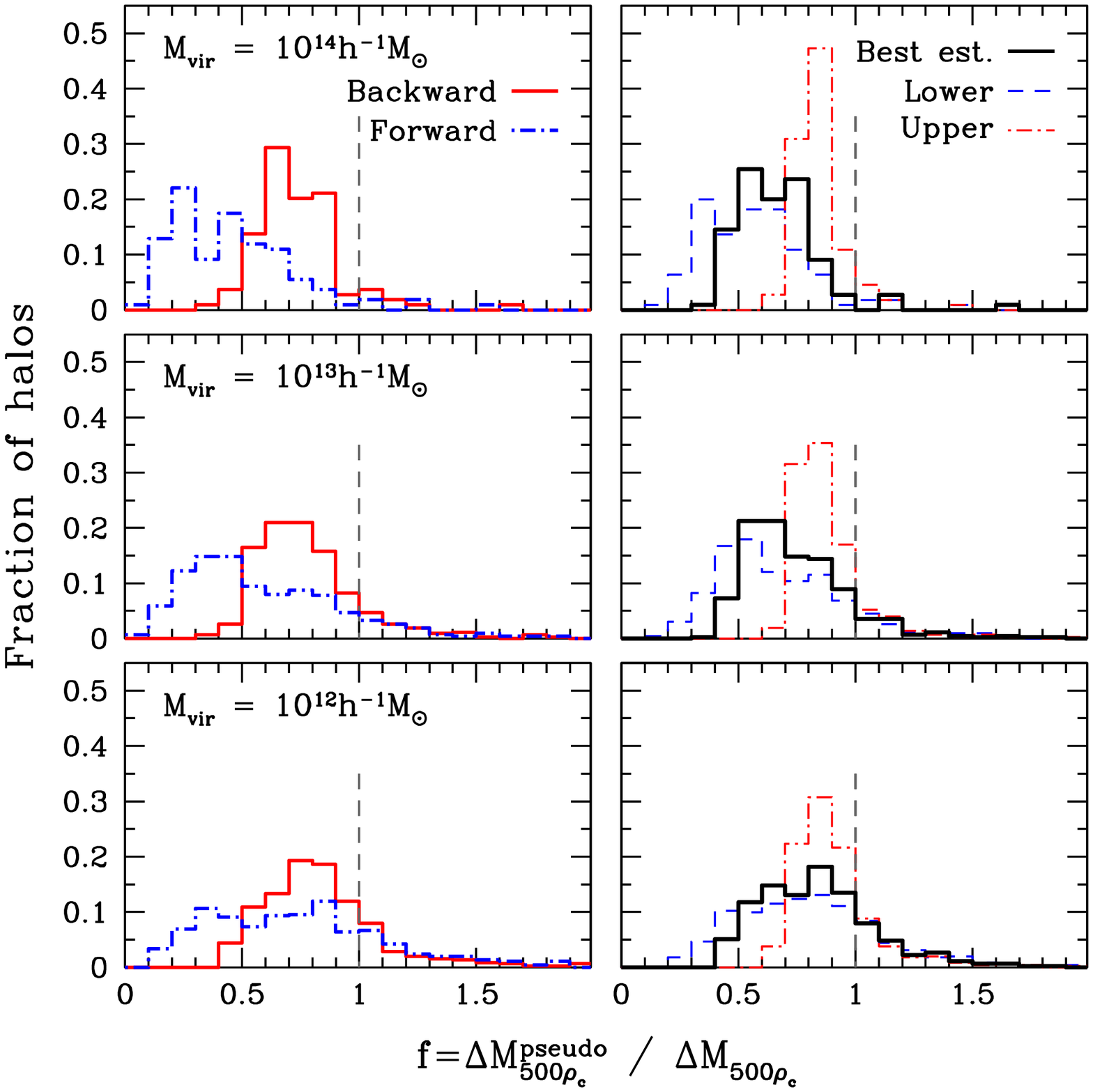} 
\caption{Histograms of the fraction of mass evolution estimated to be due to pseudo-evolution. The three rows correspond to three mass bins of $\mvir(z=0)$, $1.0 \times 10^{14} < M < 1.5 \times 10^{14} \msunh$ (110 halos, top row), $1.0 \times 10^{13} < M < 1.1 \times 10^{13} \msunh$ (502 halos, center row), and $1.00 \times 10^{12} < M < 1.01 \times 10^{12} \msunh$ (452 halos, bottom row). Note that we bin halos using masses defined with respect to the virial overdensity $\Delta_{\rm vir}\approx 358$ (the average mass of each bin is indicated in the legend), while we study the mass evolution defined with respect to overdensities $200\bar{\rho}$ (the left two columns) and $500\rho_c$ (the right two columns). For each mass definition, the left column shows the fraction of the total mass evolution due to pseudo-evolution estimated from the forward (blue) and backward (red) evolution. The second and fourth columns show the lower and upper bounds on pseudo-evolution, as well as the best estimate (see Section \ref{subsec:nonstatic} for the definition of these estimators). For a truly static density profile, all estimators give $f_{\rm pseudo} = 1$ (indicated by the gray vertical lines).}
\label{fig:individual}
\end{figure*}

In Section \ref{subsec:mah200b}, we explored what fraction of halo mass evolution is due to pseudo-evolution {\it on average}. In this section, we examine mass evolution of {\it individual} halos, using $z=1$ main progenitors of the $z=0$ halos identified as described in Section \ref{subsec:sim} above.  For each halo, we compared the difference in mass due to pseudo-evolution, $\Delta M_{\rm pseudo}$, and the actual difference in the virial mass $\Delta M$ of the progenitor and descendant halos. Figure \ref{fig:individual} shows histograms of this fraction for two different mass definitions, $\mmean$ (left two columns) and $\mcrit$ (right two columns). For each mass definition, the left column shows forward and backward evolution, and the right column the lower and upper limits, as well as the best estimator as described in Section \ref{subsec:nonstatic}.

First, let us consider the general meaning of the fraction of pseudo-evolution, $f_{\rm pseudo}$. If $0<f_{\rm pseudo}<1$,  the halo mass growth was due to both pseudo-evolution and physical accretion. Figure \ref{fig:individual} shows that this is the case for most halos. The case of $f_{\rm pseudo}=1$ corresponds to pure pseudo-evolution, while $f_{\rm pseudo}>1$ indicates that the density profile of the halo has decreased since $z=1$. These halos may have undergone tidal stripping, even though they are located outside formal virial radius of any halo at $z=0$.  Examples of these cases are explicitly discussed in Section \ref{subsec:nonevo}. 

The results in the first and third columns of Figure \ref{fig:individual} are in agreement with the average results in Figures \ref{fig:mah200b} and \ref{fig:mah500c}. Forward evolution predicts a lower value of pseudo-evolution than backward evolution, and exhibits somewhat larger scatter. The difference is particularly large for cluster-sized halos, as expected from the discussion in Section \ref{subsec:nonstatic}. The differences between the pseudo-evolution in $\mmean$ and $\mcrit$ are relatively insignificant, in agreement with Figure \ref{fig:mah500c}. As we discussed above, for halos undergoing a real mass increase due to accretion and merging, forward evolution estimate provides a lower limit estimate of the amount of pseudo-evolution. However, given that backward evolution can either underestimate or overestimate it, we cannot put an upper bound on the amount of mass pseudo-evolution (except, of course, the useless upper bound that all of the mass evolution is due to pseudo-evolution).

This issue is alleviated if we use density profiles at both $z_\rmi$ and $z_\rmf$. The second and fourth columns of Figure \ref{fig:individual} show the lower and upper bounds, as well as the corresponding best estimates of pseudo-evolution. As one could expect, the distributions for the lower bound are fairly similar to those of forward evolution, and the upper bound mimics backward evolution. This indicates that backward and forward evolution do, in general, bracket the true amount of pseudo-evolution. Note that for physically decreasing density profiles ($f_{\rm pseudo}>1$), the lower and upper bounds are reversed. This reversal is visible in the $M = 10^{12} \msunh$ sample. It is important to keep in mind that the best estimate was based on the assumption that the density profile grows or decreases uniformly at all radii, and linearly with cosmic time. This is certainly not the case for all halos, and the best estimate should be interpreted as the best guess of the true amount of pseudo-evolution.

Nevertheless, the distributions shown in Figure \ref{fig:individual} highlight the importance of pseudo-evolution over a wide range of halo masses, confirming the average trends discussed in the previous section.  For halos in the lowest mass bin, pseudo-evolution dominates over physical accretion, although the scatter in $f_{\rm pseudo}$ is large. For halos in the largest mass bin, the best estimate distribution is peaked around $f_{\rm pseudo} \approx 0.7$, consistent with Figure \ref{fig:mah200b}. However, the best estimate does not include the effects of mergers which might be the dominant source of growth for cluster-sized halos. Thus, the lower bound might be a more sensible estimate to consider for large halos. The lower bound estimate allows for most of the mass evolution to be due to physical growth, but there is still a significant populations of halos for which $f_{\rm pseudo} > 0.5$. This leads to the somewhat surprising conclusion is that even for halos with $\mvir = 10^{14} \msunh$, pseudo-evolution can account for almost the entire mass evolution since $z=1$. Thus,  individual halo mass evolution histories may exhibit mostly pseudo-evolution, even for halo masses for which the pseudo-evolution contributes a small fraction of mass evolution on average. Our results are consistent with a recent study by \citet{wu_13_rhapsody1}  who find that for cluster-sized halos with the average mass of $\mvir=10^{14.8}M_{\odot}$ the quartile of halos with the highest formation redshift experiences almost no physical accretion after $z=1$, and that its mass evolution is almost entirely accounted for by pseudo-evolution. Thus, even for clusters, commonly assumed to be dynamically young and still actively growing systems, as much as a quarter of the population may have experienced little physical mass accretion during the last seven billion years.

\subsection{The Mass (Non-) Evolution of Low-mass Halos}
\label{subsec:nonevo}

\begin{figure*}
\centering
\includegraphics[trim = 0mm 5mm 58mm 10mm, clip, scale=0.62]{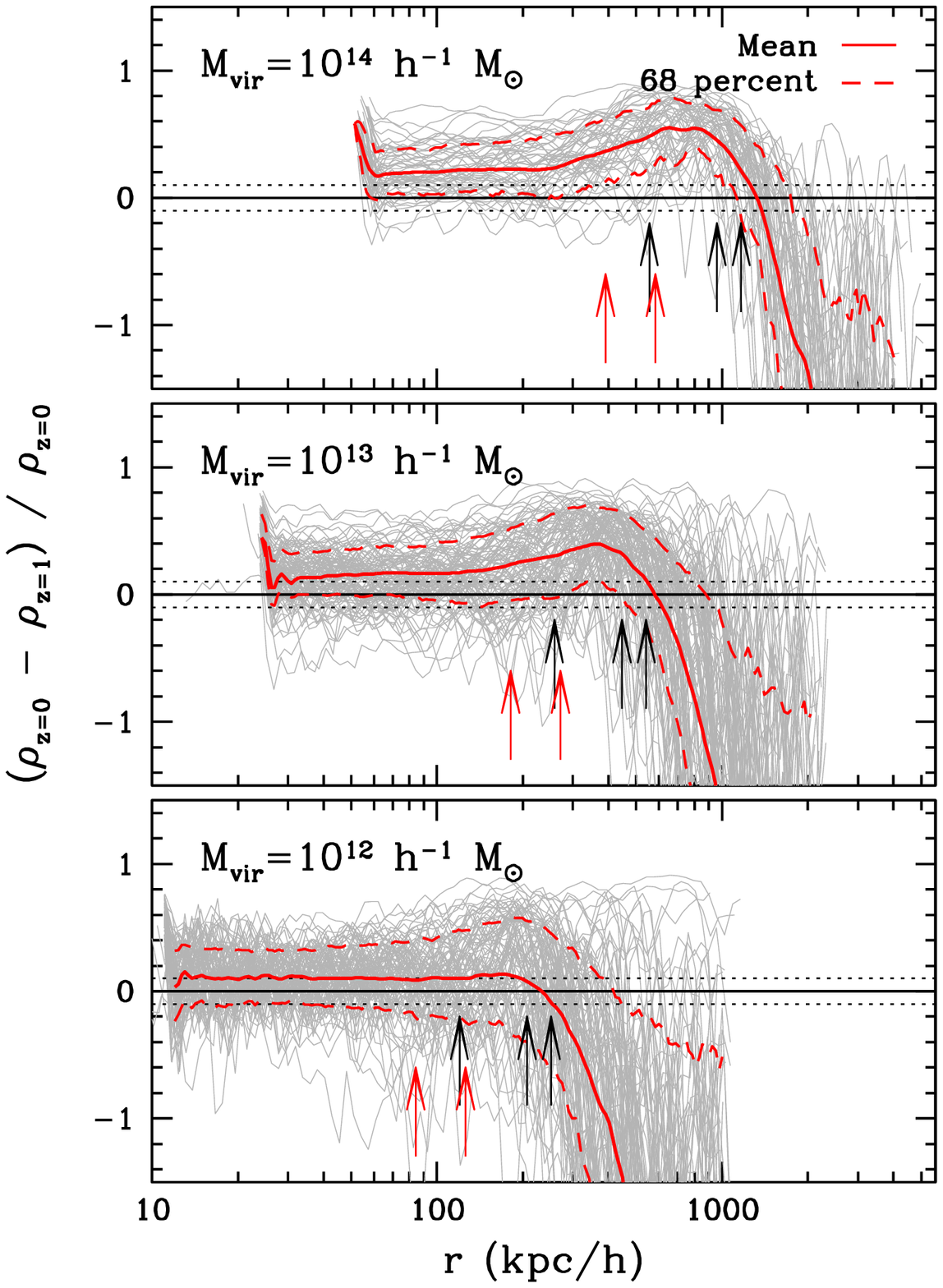} 
\includegraphics[trim = 0mm 5mm 65mm 10mm, clip, scale=0.62]{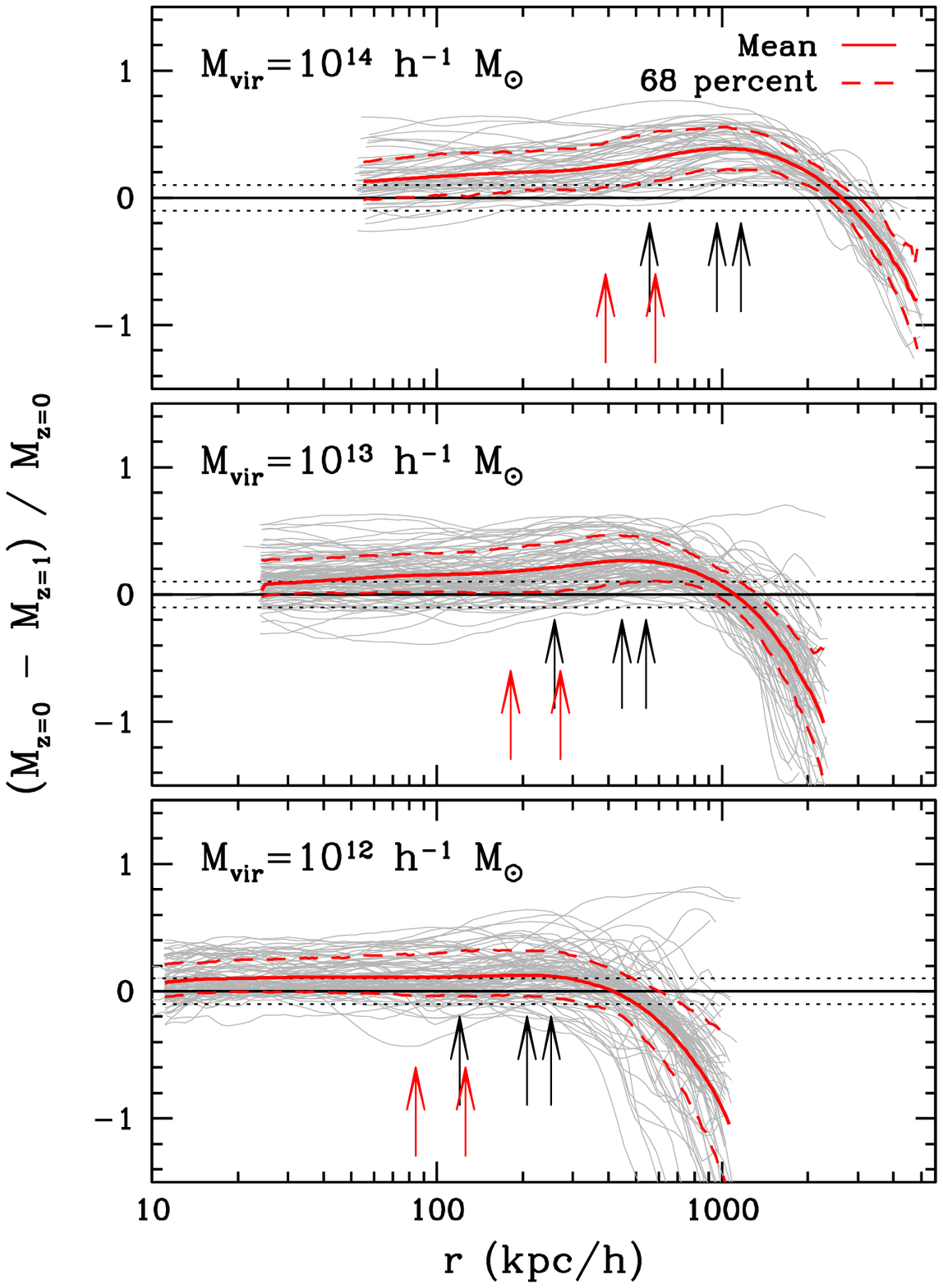} 
\caption{Difference between the density (left column) and enclosed mass (right column) profiles between $z=1$ and $z=0$, for the same mass bins as in Figure \ref{fig:individual}. For each mass bin, gray lines show the difference for about 100 individual halos, while the mean is plotted in red and the dashed lines indicate the range containing 68\% of halos. The arrows indicate, from left to right, $\rcrit$, $\rvir$ and $\rmean$, with $z=0$ in black and $z=1$ in red. Note that at $z=1$, $\rvir$ and $\rmean$ happen to overlap almost exactly. The black line at 0 and the dashed lines at $\pm 10\%$ are intended to guide the eye. Pseudo-evolution is caused by the shift in the halo boundary (from red to black arrows) due to evolving reference density. This plot demonstrates that the physical density profiles of low-mass halos grow by only $\approx 10\%$ on average between $z=0$ and $z=1$. This is true for both the differential density and enclosed mass profiles, though the density exhibits a much larger scatter.}
\label{fig:directcomp}
\end{figure*}

One of the most striking consequences of the results presented in Figures \ref{fig:mah200b} and \ref{fig:mah500c} is that the physical density profiles of most low-mass halos ($M_{200\bar{\rho}} \lsim 10^{12}\msunh$), and even some cluster halos, barely change after $z=1$. We seek to demonstrate this directly in Figure \ref{fig:directcomp} which shows the evolution of density profiles and enclosed mass profiles between $z=1$ and $z=0$ for the same mass bins as in Figure \ref{fig:individual}. Though density and enclosed mass are obviously connected, fluctuations in the density profile are often smoothed out in the enclosed mass profile, and it is important to consider both. As expected from our previous results, there is significant scatter in the evolution of individual halo density profiles (shown with gray lines), and much smaller scatter in the enclosed mass profiles. However, the mean evolutions show some very clear trends. For low-mass halos, both density and enclosed mass have grown by 10\% between $z=1$ and $z=0$, regardless of the overdensity used (and thus the virial radius). At the outskirts, the density profiles show a sharp decrease on average, starting at about $\rvir(z=0)$. This result confirms  findings of \citet[][see their Figure 16]{cuesta_08_infall}. The decrease manifests itself in the enclosed mass profiles, but at significantly larger radii, and has thus little effect on the evolution of the SO mass. The observed growth of $\approx 10\%$ is in excellent agreement with the results in Figure \ref{fig:mah200b}. For larger halos, three distinct contributions to mass growth become apparent: the actual evolution of the density profile at the virial radius, the increase in radius (pseudo-evolution), and the particular increase in enclosed mass between the old and new virial radii (see Section \ref{subsec:nonstatic} for a mathematical description of these contributions). Given that the right column of Figure \ref{fig:directcomp} shows the difference in enclosed mass rather than its absolute value, it is not easy to estimate the difference in mass contributed by pseudo-evolution.  

The non-evolution of halo density profiles demonstrated in Figure \ref{fig:directcomp} implies that low-mass halos undergo only a small amount of physical accretion since $z=1$.  We investigated the amount of physical infall into halos by extracting profiles of the average radial velocity within radial bins, similar to the density profiles, from the Bolshoi simulation. We confirmed the earlier result of \citet{cuesta_08_infall} that the average infall velocity in low-mass halos only amounts to a small fraction of the circular velocity at $\rmean$, $v_{200}$. Furthermore, we estimated the total physical accretion since $z=1$ by assuming that the infall profile remains static. This naive estimate is in agreement with our previous results, showing that for low-mass halos, the infall profiles at $z=1$ suggest physical accretion to grow the halo by less than 20\% within the virial radius at $z=0$. For cluster-sized halos, the accretion estimate gives factors of a few times the halo's mass at $z=1$.

In summary, all our measurements point to a coherent picture, where low-mass halos grow predominantly through pseudo-evolution after $z=1$, and encounter very little actual physical growth. Though the change in density profiles is subject to significant scatter, the observed non-evolution in density lends credibility to our initial estimate using static density profiles as a first-order approximation of pseudo-evolution.

\subsection{Halo Mass Function from the Static Halo Model}
\label{subsec:mass_func}

\begin{figure*}
\centering
\includegraphics[trim = 0mm 5mm 10mm 75mm, clip, scale=0.7]{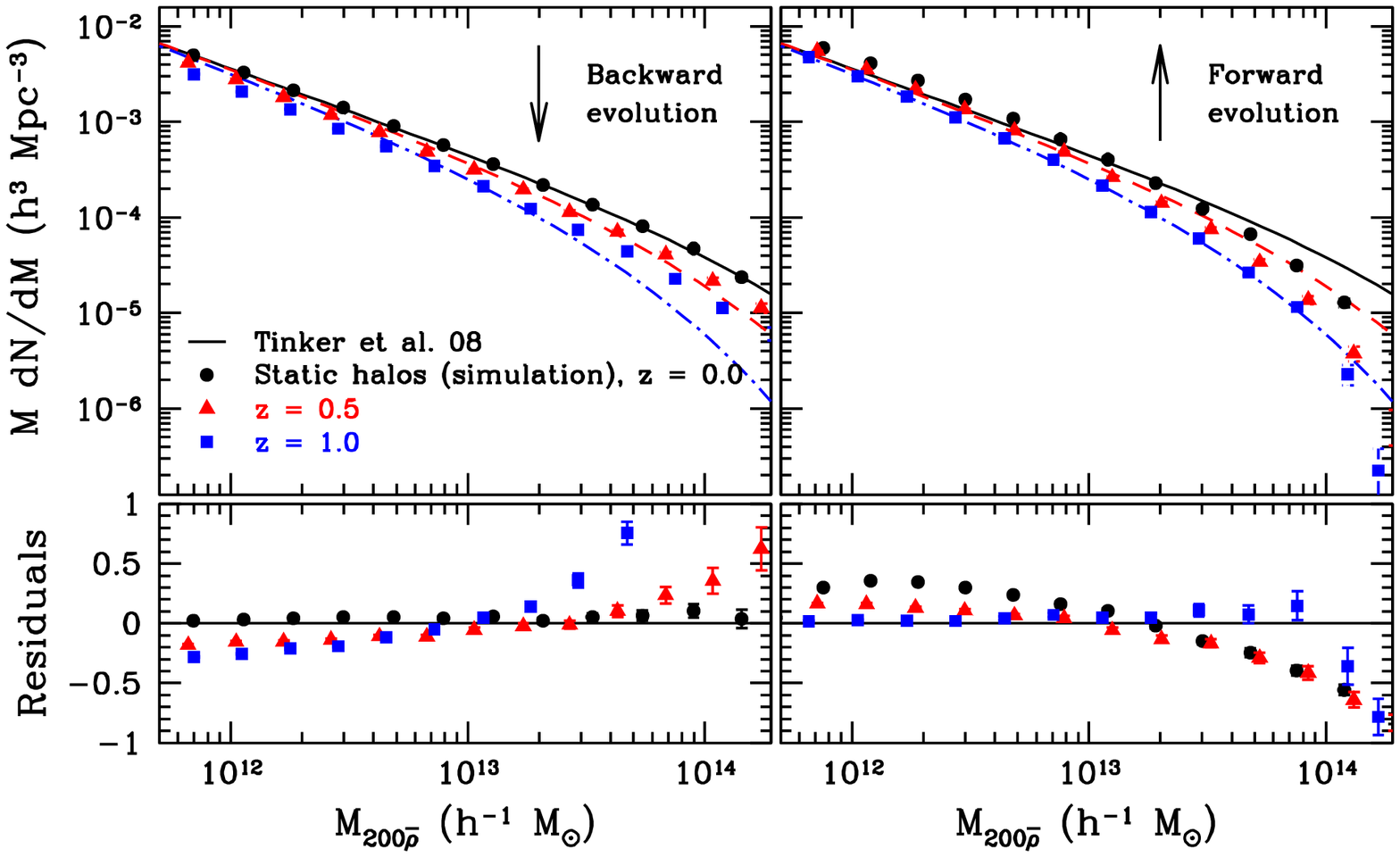} 
\caption{Comparisons of the mass function evolution due to pseudo-evolution (points) to the actual evolution of the mass function (lines), as quantified by \citet{tinker_08_mass_func}. The data points show the pseudo-evolution as derived from the density profiles of the Bolshoi halos, evolving backward in time (left panel) and forward in time (right panel). The error bars indicate the Poisson uncertainty. The bottom panels show the fractional difference between the points and the T08 mass function for the corresponding redshifts. The static halo model predicts that the mass function changes via a simple, uniform shift to higher mass, which is significantly different from the actual evolution of halo mass function. This difference indicates that in addition to mass evolution, a substantial fraction of low-mass halos disappear as they are incorporated into the virial radii of larger halos. See Section \ref{subsec:mass_func} for a detailed discussion.}
\label{fig:mass_func_tinker}
\end{figure*}

Given the success of the static halo model in reproducing mass evolution histories for low-mass halos, we would like to investigate the impact of the mass definition on the evolution of the halo mass function. For this purpose, we start from the $z=0$ halo mass function calibration of \citet[][hereafter T08]{tinker_08_mass_func}, and use the mass evolution history (backward evolution) inferred from our static halo model to predict the resulting evolution of the halo mass function. In the static halo model, the mass assigned to density peaks becomes smaller as redshift increases, which results in a mass-dependent shift of the halo mass function ($M dN/dM $) toward the left. In order to quantify this effect, we used the same procedure as for the halo mass evolution. We selected the density profiles of halos of mass $\mvir >2\times10^{11}\msunh$ extracted from the simulation and calculated the expected evolution of mass assuming that the density profiles around peaks stay constant in physical units.

Before comparing these results to the {\it actual} physical evolution of the mass function observed in numerical simulations, we first establish that these results match the analytical prediction of Equation (\ref{eq:mevol}). Given that the corresponding mean mass evolution histories agree to a few percent (Figure \ref{fig:mah200b}), we naively expect good agreement between the mass functions as well. However, Figure \ref{fig:mah200b} also reveals significant scatter in the pseudo-evolution of simulated halos, which could cause disagreement with the analytically predicted mass function. 

We find that in the case of the {\it backward} evolution of $\mmean$, the prediction of the static halo model agrees well (to better than $5-10\%$) with the expected evolution from the actual profiles of the halos at $z=0$. For the case of mass definitions using higher overdensities, such as $\mcrit$, we expect even better agreement, because the static halo describes the pseudo-evolution of halos in simulations more accurately (see Figure \ref{fig:mah500c}). For the prediction of the static halo model for the case of {\it forward} evolution, we start from the halo mass function of T08 at $z=1$, and evolve forward in time to $z=0$. We find similarly good agreement between this prediction and the mass function predicted when pseudo-evolving the simulated density profiles forward in time. 

Note that there is a discrepancy of $\sim 5\%$ between the $z=0$ and $z=1$ calibration of the T08 mass function and the mass function obtained from the Bolshoi simulation. Given this initial offset, we cannot expect a smaller discrepancy at subsequent redshifts.
Furthermore, we considered the impact of statistical bias due to the presence of scatter in the mass evolution histories \citep{eddington_13_stat_bias}. It is evident from Figure \ref{fig:mah200b} that there is a non-negligible scatter in the mass evolution histories, and that this scatter is somewhat larger in the forward evolution case than in the backward evolution case. As the number density of halos is a decreasing function of halo mass, the number of halos that get up-scattered into a particular mass bin can be larger than the number of halos that get down-scattered out of that mass bin. However, we found that this bias does not influence the results appreciably. 

\subsection{Comparison with the True Mass Function}
\label{subsec:mass_func_true}

Having convinced ourselves that the halo mass functions predicted by the static halo model and halo profiles are consistent, we now wish to investigate the impact of pseudo-evolution on the {\it true} evolution of the halo mass function. We use the calibration of the mass function provided by T08 to reflect the true evolution of the mass function measured in simulations. In Figure \ref{fig:mass_func_tinker}, we present the comparison of the T08 mass function at three different redshifts with the evolution of the mass function due to the pseudo-evolution of density profiles from the simulation. Let us first focus on the left hand panel which shows the {\it backward evolution} case. Given that our estimates of the mass evolution histories of low mass halos matched those observed in simulations (Figure \ref{fig:mah200b}), we expect good agreement with the mass function at the low-mass end, and discrepancies predominantly at the high-mass end. The left hand panel of Figure \ref{fig:mass_func_tinker}, however, reveals significant disagreement at both mass ends. Because the cause of deviations are different at the low and high-mass ends, we shall discuss those regions separately.

At the high-mass end, we have shown that the growth of high-mass halos is largely due to physical accretion rather than pseudo-evolution. This rapid growth implies that the T08 mass function decreases strongly with redshift. As we evolve backwards in time, the progenitors of high-mass halos would need to be more massive than in reality if the growth was solely due to changing mass definition. The static halo model, therefore, overpredicts the number of large halos at $z=1$.

At the low-mass end, computing only the pseudo-evolution underpredicts the T08 calibration at $z=0.5$ and $1$ by roughly $20-30\%$ (left panel of Figure \ref{fig:mass_func_tinker}). From the calibration of T08, it appears that the number density of low mass halos ($M \lsim 2 \times 10^{12} \msunh$) stays constant since $z=1$. The common explanation for this observed non-evolution is that low-mass halos have already collapsed and do not physically grow in mass or number. However, it is clear from Figure \ref{fig:mah200b} that these halos do indeed undergo a significant mass evolution just due to pseudo-evolution. This conflict can be resolved by noting that, in the backward evolution case, the radii of the halos reduce as we evolve the masses to higher redshifts. This may uncover substructures in the outer part of the halos, which can potentially be counted as isolated halos at those higher redshifts. Thus, the non-evolution of the halo mass function at the low mass end must be a result of the fortuitous cancellation of the effect of mass evolution and the addition of low mass halos at the outskirts of bigger halos. As we limited our analysis to use only isolated halos at $z=0$, we did not quantify this effect in the case of backward evolution. 

However, we can investigate this effect by taking the static density profiles at $z=1$ and evolving them forward to $z=0.5$ and $0$ ({\it forward evolution}). In this case, the low-mass halos in the outskirts of larger mass halos should get absorbed. When we calculate the masses of halos by using static density profiles, we partially account for this effect by removing small mass halos whose centers end up within the radius of larger mass halos at $z=0.5$ and $0$ as discussed in Section \ref{subsec:mah200b}. The right hand panel of Figure \ref{fig:mass_func_tinker} shows the comparison between the forward evolution of the mass function of halos from the Bolshoi simulation, and the T08 mass function. The comparison shows that the discrepancy at the low mass end still persists, even after removing low-mass subhalos. The reason for this discrepancy is our implicit assumption that these density peaks are stationary. While we remove some subhalos as they are absorbed into larger halos, more halos which appear isolated at $z=1$ would suffer the same fate if we took their infall motion toward larger objects into account.  This is consistent with our observation that high-mass halos (which subsume the lower mass halos in their outskirts) undergo some physical accretion in addition to the pseudo-evolution.

A naive comparison of the mass function differences between the forward and the backward evolution case at the low-mass end (see the bottom panels of Figure \ref{fig:mass_func_tinker}) seems to suggest that the effect of removing subhalos is very small. However, we note that for the case of forward evolution, it is extremely important to remove such subhalos. Due to their proximity to a larger host halo, the outskirts of their density profiles contain significant mass contributions from the host halo. As the virial radius increases toward lower redshift, such subhalos gain a significant fraction of the host halo mass. If they are not removed, this unphysical mass evolution results in a large {\it scatter} in the mass evolution histories of the low mass halos. This scatter can have a large effect on the estimated mass function due to Eddington bias (see Section \ref{subsec:mah200b}). Thus, {\it not} removing the 14\% of halos which become subhalos by $z=0$ can lead to residuals of over 100\% when comparing to the T08 mass function for the case of forward evolution.

%%%%%%%%%%%%%%%%%%%%%%%%%%%%%%%%%%%%%%%%%%%%%%%%%%%%%%%%%%%%%%%%%%%%%%%%%%
% DISCUSSION
%%%%%%%%%%%%%%%%%%%%%%%%%%%%%%%%%%%%%%%%%%%%%%%%%%%%%%%%%%%%%%%%%%%%%%%%%%

\section{Discussion}
\label{sec:discussion}

In the past two sections, we have demonstrated that pseudo-evolution due to changing reference density has a significant impact on the overall evolution of SO mass (often called {\it mass accretion history}). In this section, we expand on some of the implications of this result for our understanding of the scaling relations between various observables and halo mass. This includes the concentration-mass relation, the relation between stellar content and halo mass, and scaling relations for galaxy clusters.

\subsection{The Concentration-Mass Relation}

\begin{figure}
\centering
\includegraphics[trim = 0mm 5mm 90mm 5mm, clip, scale=0.6]{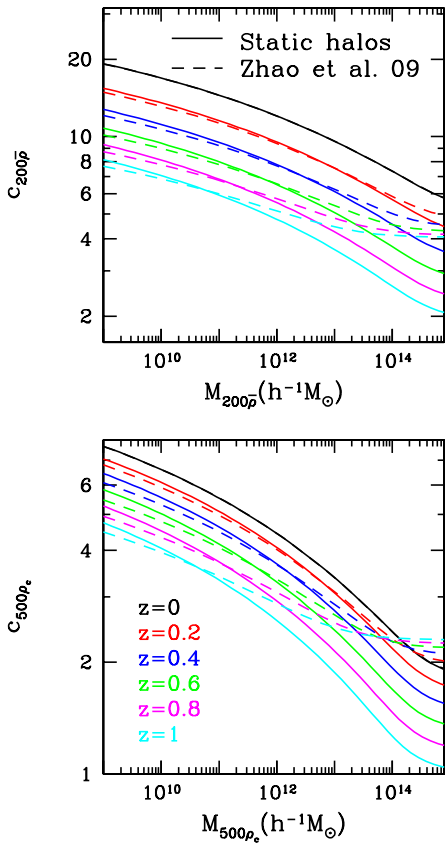} 
\caption{Evolution of the concentration-mass relation with redshift expected if the physical density around density peaks is unchanged over time. The solid lines show the analytical prediction of Equations (\ref{eq:cevol})-(\ref{eq:mevol}). From top to bottom, the lines correspond to redshifts $0-1$ in steps of $0.2$. The dashed lines show the concentration-mass relation as a function of redshift from the physical model of Z09. This figure demonstrates that at low masses a large fraction of the observed evolution in the $c-M$ relation is simply due to pseudo-evolution.}
\label{fig:cm_relation}
\end{figure}

In the presence of pseudo-evolution, the virial radius of a halo grows with time, even though the halo's physical density profile (and thus its scale radius, $r_s$) remain unchanged. However, because the virial radius {\it does} change due to pseudo-evolution, the concentration, $c$, grows at the same rate as the virial radius. Thus, we expect a significant evolution in the concentration-mass relation (hereafter $c-M$ relation), even if the physical density profile of a halo remains unchanged. 

We use the static halo model to estimate the magnitude of this evolution. The prediction of the static halo model for two mass definitions is shown with solid lines in  Figure \ref{fig:cm_relation}, and is in qualitative agreement with the evolution observed in numerical simulations such that the concentration of halos of a given mass decreases with increasing redshift. In reality, we expect that halos undergo some true physical evolution of mass due to accretion and merging, especially at the high-mass end, which will result in quantitative discrepancies between the {\it true} evolution of the $c-M$ relation observed in numerical simulations and our static halo model predictions. 

The dashed lines in Figure \ref{fig:cm_relation} show the redshift-dependent $c-M$ relation obtained from the models of Z09 which have been calibrated to reproduce the evolution of this relation in numerical simulations. The comparison clearly shows that most of the evolution in the $c-M$ relation at the low-mass end can be accounted for by pseudo-evolution of halo radius at different redshifts. For high-mass halos, however, the Z09 model captures the $c-M$ evolution due to their significant physical mass accretion, and is thus not completely reproduced by our model of pseudo-evolution. Note that the static halo model quantitatively reproduces the evolution of the $c-M$ relation for low-mass halos obtained by \citet{bullock_01_halo_profiles}. 

\subsection{Implications for Galaxy Formation}

The total stellar content (or the stellar light) that we observe as galaxies in a halo is the integral result of the complex interplay between a variety of processes such as star formation, feedback from young stars, SNe and supermassive black holes, and galactic outflows, all of which occur within dark matter halos. The redshift evolution of the scaling relation between this stellar mass (or luminosity) and halo mass can provide important observational clues regarding these different physical processes, in particular their efficiency as a function of halo mass. A number of studies have investigated the scaling relations between stellar mass and the mass of the halo  they inhabit, and how these scaling relations change with time \citep{conroy_07_sk, brown_08_hod, behroozi_10_shmr,moster_10_shmr, abbas_10_vvds, wake_11_newfirm, leauthaud_12_shmr, yang_12_dm_galaxy}. However, when connecting the observed evolution of the scaling relations to the underlying physics, it is crucial to account for the pseudo-evolution of halo mass. 

One of the striking implications of our results is that almost all of the mass evolution of most galactic-sized halos ($\mmean[z=0]\simlt10^{12}\msunh$) since $z=1$ can be attributed to pseudo-evolution (see Figure \ref{fig:individual}). The density profiles around the peaks of such halos have stayed static and not evolved physically \citep[see also][]{prada_06_outerregions, diemand_07_haloevolution, cuesta_08_infall}. Even for cluster-sized halos, as much as a quarter of the population do not experience appreciable physical accretion between $z=1$ and $z=0$ \citep[see also][]{wu_13_rhapsody1}.

Given that for galaxy-sized halos the physical accretion plays a minor role compared to pseudo-evolution, the impact of pseudo-evolution must be considered while interpreting the evolution of scaling relations and relating them to the underlying physical processes. For example, the ratio of stellar mass to halo mass (SHMR), and its evolution with redshift, gives a quantitative measure of how the star formation efficiency in a halo of given mass evolves with redshift. The peak of the star formation efficiency lies at roughly $\mmean[z=0]\simeq10^{12}\msunh$, and has been observed to shift to higher values from $z=0$ to $1$ \citep{moster_10_shmr, behroozi_10_shmr, leauthaud_12_shmr}. However, equal halo masses at two different redshifts correspond to different physical density peaks due to pseudo-evolution. Thus, the rate at which the similar density peaks become inefficient can differ from the estimates at fixed halo mass.

As a second example, let us consider the evolution of the SHMR at the more massive end. The stellar mass in such halos is dominated by satellite galaxies. Pseudo-evolution will lead to a constant fractional increase in both the stellar content and the halo mass if the distribution of satellite galaxies in and around halos, to first order, follows the matter density distribution, and if there is no radial segregation in the stellar mass of satellite galaxies. In this case, stellar and halo mass grow by the same factor, and the SHMR at the massive end undergoes an evolution with redshift which is qualitatively very similar to the evolution observed by \citet{leauthaud_12_shmr}. We will perform a quantitative comparison between the evolution of the SHMR due to pseudo-evolution of halo mass and the observed SHMR evolution in future work.

\subsection{Implications for Cluster Scaling Relations}

As discussed in Section \ref{sec:intro}, scaling relations between the baryonic properties of clusters, such as X-ray temperature, gas mass, or entropy, and the mass of the cluster's dark matter halo, are key to our understanding of clusters and their use in cosmology. The simplest model for these scaling relations relies on the assumption that cluster halos collapse in a self-similar fashion \citep{kaiser_86_clusters}. In this model, the temperature $T$, for example, scales with halo mass as
\begin{equation}
T \propto \frac{M}{R}
\end{equation}
where $M$ denotes the mass within the radius $R$, and $T$ is measured at $R$ (KB12) . If the halo mass is defined as a spherical overdensity mass (mass definitions such as $\mcrit$ or $M_{2500\rho_\rmc}$ are commonly used for clusters), the above scaling relation can be expressed as 
\begin{equation}
\label{eq:scalingrel2}
T \propto (\Delta_\rmc \rho_\rmc)^{1/3} M_{\Delta}^{2/3}.
\end{equation}
Noting that $\Delta_\rmc$ is a constant, but that the critical density evolves with $E^2(z)$, the evolution with redshift can be incorporated into the scaling relation as
\begin{equation}
\label{eq:mt}
T \propto \left [ E(z) M_{\Delta} \right]^{2/3}.
\end{equation}
Unfortunately, the $E^2(z)$ factor only accounts for the evolution of $R$ (due to the evolution of the $\rho_\rmc$ factor in Equation (\ref{eq:scalingrel2})), but {\it not} the pseudo-evolution of $M_{\Delta}$. Thus, for a halo whose density profile remains constant, the scaling relation predicts that the temperature will increase with time without any particular physical reason.

As shown in Figure \ref{fig:mah500c}, the mass evolution of $\mcrit$ since $z=1$ is only partly due to pseudo-evolution, but also contains a large contribution from actual accretion. Nevertheless, pseudo-evolution accounts for a factor of a third. If this contribution is not taken into account, pseudo-evolution could masquerade as a deviation from self-similar behavior. Interestingly, the pseudo-evolution is not apparent in the actual evolution of cluster scaling relations, such as the $M-T$ relation \citep[see, e.g.,][]{nagai_06, stanek_10_scaling}. This may be because the $E(z)$ factor in Equation (\ref{eq:mt}) fortuitously compensates for most of the pseudo-evolution. We intend to further investigate the impact pseudo-evolution has on scaling relations in a future study. However, finding a mass definition which allows us to disentangle the effects of pseudo-evolution on cluster scaling relations is certainly a challenging task, especially because different cluster observables have different dependencies on the exact boundary used to define a halo.

%%%%%%%%%%%%%%%%%%%%%%%%%%%%%%%%%%%%%%%%%%%%%%%%%%%%%%%%%%%%%%%%%%%%%%%%%%
% CONCLUSION
%%%%%%%%%%%%%%%%%%%%%%%%%%%%%%%%%%%%%%%%%%%%%%%%%%%%%%%%%%%%%%%%%%%%%%%%%%

\section{Conclusions}
\label{sec:conclusion}

Several authors have pointed out that SO masses undergo an evolution due to the changing reference density \citep{diemand_07_haloevolution, cuesta_08_infall, kravtsov_12_cluster_review}. In this paper, we have studied  this spurious {\it pseudo-evolution} of mass quantitatively. Our main results and conclusions are as follows. 

\begin{enumerate}
\item We have demonstrated that for all halo masses a significant fraction of the halo mass growth since $z=1$ is due to pseudo-evolution rather than the actual physical accretion of matter. For the majority of low-mass halos ($\lsim 10^{12} \msunh$), pseudo-evolution accounts for almost all of the evolution in mass since $z=1$. Even for a fraction of large cluster halos pseudo-evolution represents the dominant mode of mass growth.
\item We found the scatter in the amount of pseudo-evolution to be significant for all halo masses, corresponding to significant scatter in the physical evolution of halo density profiles.
\item We have presented a mathematical definition of physical evolution and pseudo-evolution, and have shown that the exact amount of pseudo-evolution can only be computed using many snapshots which are finely spaced in time.
\item We have shown that a simple analytical model based on the assumption of static NFW profiles reproduces the average pseudo-evolution observed in the Bolshoi simulation to a few percent accuracy.
\item We directly demonstrated that the physical growth in density profiles since $z=1$ is about 10\% on average for galaxy-sized halos. This finding was supported by the infall velocities around low-mass halos at $z=1$ which are insufficient to facilitate significant growth.
\item We investigated the effect of pseudo-evolution on the halo mass function $dN/d \ln(M)$, and found it to simply shift the function toward higher masses. We propose that the non-evolution of the mass function at low masses since $z=1$ constitutes a fortuitous cancellation between pseudo-evolution and the absorption of small halos into larger halos.
\end{enumerate}

We have left some questions for future investigations. For example, in this paper we restricted ourselves to halos with SO definitions. However, another popular way to identify and define halos is to use the friends-of-friends (FoF) algorithm, which relies on a linking length (which is fixed relative to the average inter-particle comoving separation) rather than an overdensity to define masses. It is well known that the density of FoF halos at their boundary depends upon linking length \citep[see e.g.][]{frenk_88_haloformation, lukic_09_halostructure, more_11_fof}. Given that the linking length parameter is constant in comoving coordinates, its physical length increases with time as the scale factor $(a=[1+z]^{-1})$. This implies that for a static halo density profile, the extent of the FoF halo will increase with time, leading to pseudo-evolution. For example, in a study of the mass evolution history of the FoF halos, \citet{fakhouri_10_halogrowth} disentangle the growth into accretion of resolved halos and the accretion of a diffuse component. It is clear that the diffuse component will include a contribution from pseudo-evolution, thus overestimating the fraction of actually accreted diffuse matter compared to the fraction accreted from merging halos. Therefore, the pseudo-evolution of the FoF halos will need to be carefully investigated.

The eventual goal of such investigations will be to find a practical mass definition that can be used both in analyses of simulations and observations and which does not suffer from pseudo-evolution, thereby allowing for a more robust formulation of scaling relation evolution. We hope to address this question in future work.

%%%%%%%%%%%%%%%%%%%%%%%%%%%%%%%%%%%%%%%%%%%%%%%%%%%%%%%%%%%%%%%%%%%%%%%%%%
% ACKNOWLEDGMENTS
%%%%%%%%%%%%%%%%%%%%%%%%%%%%%%%%%%%%%%%%%%%%%%%%%%%%%%%%%%%%%%%%%%%%%%%%%%

\section*{Acknowledgments}

We are indebted to Anatoly Klypin for providing us with the N-body simulation data used in this paper. We thank Frank van den Bosch, Eduardo Rozo, Matthew Becker, and Samuel Leitner for useful discussions and their comments on the draft. The MultiDark database used in this paper and the web application providing online access to it were constructed as part of the activities of the German Astrophysical Virtual Observatory as result of a collaboration between the Leibniz-Institute for Astrophysics Potsdam (AIP) and the Spanish MultiDark Consolider Project CSD2009-00064. The Bolshoi and MultiDark simulations were run on the NASA's Pleiades supercomputer at the NASA Ames Research Center.

%%%%%%%%%%%%%%%%%%%%%%%%%%%%%%%%%%%%%%%%%%%%%%%%%%%%%%%%%%%%%%%%%%%%%%%%%%
% BIBLIOGRAPHY
%%%%%%%%%%%%%%%%%%%%%%%%%%%%%%%%%%%%%%%%%%%%%%%%%%%%%%%%%%%%%%%%%%%%%%%%%%

\bibliographystyle{apj}
\bibliography{../../../_LatexInclude/sf.bib}

\end{document}

%% file: commands.tex
\newcommand{\mpch}{\>h^{-1}{\rm {Mpc}}}

\newcommand{\msun}{\>{\rm M_{\odot}}}

\newcommand{\msunh}{\>h^{-1}\rm M_\odot}

\newcommand{\kmsmpc}{\>{\rm km}\,{\rm s}^{-1}\,{\rm Mpc}^{-1}}

\def\mvir{M_{\rm vir}}
\def\rvir{R_{\rm vir}}
\def\gcm3{\mathrm{g} / \mathrm{cm}^3}

\def\gtsima{$\; \buildrel > \over \sim \;$}
\def\ltsima{$\; \buildrel < \over \sim \;$}
\def\prosima{$\; \buildrel \propto \over \sim \;$}
\def\gsim{\lower.7ex\hbox{\gtsima}}
\def\lsim{\lower.7ex\hbox{\ltsima}}
\def\simgt{\lower.7ex\hbox{\gtsima}}
\def\simlt{\lower.7ex\hbox{\ltsima}}
\def\simpr{\lower.7ex\hbox{\prosima}}

\def\rmb{{\rm b}}
\def\rmc{{\rm c}}
\def\rmd{{\rm d}}

\def\rmf{{\rm f}}

\def\rmi{{\rm i}}

\def\rmm{{\rm m}}

\def\rms{{\rm s}}

%% file: macros.tex
\def\mmean{M_{200\bar{\rho}}}
\def\rmean{R_{200\bar{\rho}}}
\def\mcrit{M_{500\rho_c}}
\def\rcrit{R_{500\rho_c}}